\documentclass[aps,reprint,twocolumn,superscriptaddress]{revtex4-1}
\pdfoutput=1
\usepackage{mathrsfs}
\usepackage{slashed}
\usepackage{amsmath}
\usepackage{amsfonts}
\usepackage{graphicx,color}
\usepackage{subcaption}
\usepackage[colorlinks,linkcolor=red,citecolor=blue]{hyperref}

\begin{document}

%\preprint{}

\title{Hamiltonian theory for Quantum Hall systems in a tilted magnetic field: robustness of activation gaps}


\author{Kang Yang}
%\thanks{}
\affiliation{ Laboratoire de Physique Th\' eorique et Hautes Energies, CNRS UMR 7589, Universit\' e Pierre et Marie Curie, 4 place Jussieu, 75252 Paris Cedex 05, France}
\affiliation{Laboratoire de Physique des Solides,  CNRS UMR
8502, Universit\' e Paris-Sud, 91405 Orsay Cedex, France}
\author{Mark Oliver Goerbig}
\affiliation{Laboratoire de Physique des Solides,  CNRS UMR
8502, Universit\' e Paris-Sud, 91405 Orsay Cedex, France}
\author{Beno\^ it Dou\c cot}
\affiliation{ Laboratoire de Physique Th\' eorique et Hautes Energies, CNRS UMR 7589, Universit\' e Pierre et Marie Curie, 4 place Jussieu, 75252 Paris Cedex 05, France}

\date{\today}

\begin{abstract}
 We use the Hamiltonian theory developed by Shankar and Murthy to study a quantum Hall system in a tilted magnetic field.
 With a finite width of the system in the $z$ direction, the parallel component of the magnetic field introduces anisotropy into the effective
 two-dimensional interactions. The effects of such anisotropy can be effectively captured by the recently proposed generalized pseudo-potentials.
 We find that the off-diagonal components of the pseudo-potentials lead to mixing of composite fermions Landau levels, which is a perturbation to the picture
 of $p$ filled Landau levels in composite-fermion theory. By changing the internal geometry of the composite fermions, such a perturbation
 can be minimized and one can find the corresponding activation gaps for different tilting angles, and we calculate the associated  optimal metric.
 Our results show that the activation gap is remarkably robust against the in-plane magnetic field in the lowest Landau level.
\end{abstract}

%\pacs{}
%\keywords{}

\maketitle


\section{Introduction}

The fractional quantum Hall effect (FQHE) is a paradigmatic platform to realize phases beyond conventional classification, i.e. phases with topological order \cite{wen2004quantum}.
Among the properties of the topological phase, one of the most important ones is that the liquid phase of FQHE supports gapped quasi-particles satisfying
anyonic statistics, which itself is the origin of the topological characteristics such as the ground-state degeneracy.
The gap of the quasi-particle excitation is related to the stability of the fractional quantum Hall (FQH) liquid.

Besides the liquid phase,
there are plenty of other compressible and incompressible candidates for quantum Hall systems. The stability of this liquid state under different perturbations
can help us to better understand the competition between different phases. Perturbing quantum Hall systems by anisotropy is an attractive approach to look for
instabilities of liquid states. Two of the easiest realizations are to add an anisotropic mass tensor or tilt the magnetic field. For a FQH liquid, how it
reacts to anisotropic perturbation remains interesting both in theory \cite{PhysRevB.87.245315,PhysRevB.95.201116} and in
experiments \cite{xia2011evidence,PhysRevLett.108.196805}. The stripe phase \cite{PhysRevB.54.1853,PhysRevB.54.5006} and the nematic
phase \cite{PhysRevB.88.125137} with broken rotation symmetry are possible candidates in such cases. The former is common in higher ($\nu\ge2$)
Landau levels (LLs) while the latter is assumed to be the rotation-symmetry broken phase after the collective excitation mode is softened.

One method to describe the FQH liquid is to use the composite fermion (CF) language \cite{PhysRevLett.63.199,PhysRevB.41.7653,PhysRevB.44.5246}.
The FQH problem is transformed into an integer quantum Hall problem of CFs. Beyond Jain's extremely successful wave-function approach \cite{PhysRevLett.63.199,PhysRevB.41.7653},
a second-quantized Hamiltonian version has been developed by
Shankar and Murthy \cite{RevModPhys.75.1101}. The excitations of quasi-particles/holes in the FQH liquid phase translate to particle excitations
in empty CF Landau levels or hole excitations in filled CF Landau levels. The activation gap for a free particle-hole pair can be computed within
straightforward analytic calculation and a good quantitative agreement is obtained in systems with finite width \cite{RevModPhys.75.1101}, which is
the case for our problem, in this scheme.

In this paper, we study the evolution of the CF activation gaps in the presence of a parallel inplane magnetic field. The finite width of the QH system allows the
electronic orbits to be tilted away from the original plane of motion and thus to introduce an
anisotropy in the effective two-dimensional (2D) motion. It turns out that in the lowest LL, the activation gap is always robust against
the parallel magnetic field, in accordance with the conclusion of a generalized pseudo-potential study \cite{PhysRevB.96.195140}.
Meanwhile, the collective excitation gap, the magneto-roton minimum, can be subject to the anisotropy, as
numerically computed in a recent preprint \cite{yang2018behavior}. A variational metric is defined to look for the best geometry of the liquid phase
perturbed by anisotropy.  When the tilting angle of the magnetic field is increased, the optimal metric starts to deviate from the flat case, suggesting
an anisotropic liquid phase is favored.

The structure of this paper is as follows. In Sec. \ref{scrwht}, we briefly review the Hamiltonian theory by Shankar and Murthy, in view of the
activation gaps. In Sec. \ref{scmrht}, we combine Haldane's geometric variational approach with the Hamiltonian theory, showing how to characterize
an anisotropic liquid state in this sense and how this metric reacts to the external anisotropy. The single-particle
excitation spectrum for a QH system in a tilted magnetic field is calculated in Sec. \ref{scaps}. There,
the activation gaps and optimal metrics are obtained for different tilting angles and thicknesses of the sample.

\section{Review of the Hamiltonian theory}\label{scrwht}

The Hamiltonian theory developed by Shankar and Murthy \cite{RevModPhys.75.1101} has been proved successful to provide a microscopic approach for FQHE problems.
Namely, this theory allows one to calculate, within standard second-quantization methods, the activation gap for a pair of well-separated quasiparticle and quasihole.

For a 2D electron moving in a perpendicular magnetic field, the kinetic energy gives rise to LL quantization, while
the Coulomb interaction is usually treated as a smaller scale compared to the LL spacing. When a LL is partially filled, inter-LL transitions can then, in a first
approximation, be
neglected and the physical properties of the system arise from electrons projected into a single LL. In such a case, the kinetic energy is frozen and the remaining degrees of
freedom are associated with
the guiding-center coordinates $\mathbf R_e=(R_{ex},R_{ey})$ of electrons. Physically, this coordinate represents the classical center of cyclotron motion, which is a
constant of motion for a homogeneous magnetic field, and its components
obey the commutation relation $[R_{ex},R_{ey}]=-il^2$, where $l=\sqrt{\hbar/eB}$ is
the magnetic length. The projected Hamiltonian merely contains the interaction part, with the density operator replaced by the projected one:
\begin{equation}
H^p=\frac{1}{2}\sum_{\mathbf q}V_{\textrm{eff}}(\mathbf q)\rho_e(\mathbf q)\rho_e(-\mathbf q),\label{eqprjH}
\end{equation}
where $\rho_e(\mathbf q)=\sum_i\exp(-i\mathbf q\cdot\mathbf R_i)$ is the projected density operator. It is constructed by keeping only the guiding-center
coordinate in the density operator while averaging the cyclotron motion over the LL wave function. The latter averaging results in a form factor that is then absorbed
into the effective interaction $V_{\textrm{eff}}(\mathbf q)$. The effective interaction also encodes other effects such as the finite width of the sample and the anisotropy
induced by the tilted magnetic field, as we will discuss in detail below.

For a fractional filling $\nu=n_{el}/n_B$, that is the ratio between the 2D electronic $n_{el}$ and flux $n_B=eB/h$ densities, there is a huge degeneracy inside each LL. This
degeneracy of the non-interacting ground state prohibits
diagrammatic approaches that treat the interaction as a perturbation to a non-degenerate reference state. In order to eliminate this degeneracy, the strategy is
to bind each electron with vortices carrying $2s$ magnetic flux quanta opposite to the external one, forming the CF \cite{PhysRevLett.63.199,PhysRevB.41.7653,PhysRevB.44.5246}.
The CFs thus feel a weaker magnetic field and the filling changes from a fractional electronic value
$\nu_e=p/(2ps+1)$ to an integer CF filling $\nu_{\textrm{CF}}=p$. The latter CF state can then be viewed as a non-degenerate reference state for perturbative treatments.
In the language of the Hamiltonian theory,
the Hilbert space is enlarged to account for the guiding-center coordinates $\mathbf R_v$ of these vortices, which satisfy the commutation relation $[R_{vx},R_{vy}]=il^2/c^2$
with $c=2ps/(2ps+1)$ characterizing the vortex charge. A canonical transformation involving the guiding-center coordinates $\mathbf R_e$ and $\mathbf R_v$ allows us then
to introduce the CF cyclotron $\boldsymbol \eta$ and guiding-center coordinates $\mathbf R$,
\begin{equation}
\boldsymbol \eta=\frac{c}{1-c^2}(\mathbf R_v-\mathbf R_e),\ \mathbf R=\frac{\mathbf R_e-c^2\mathbf R_v}{1-c^2},\label{eqrlcfe}
\end{equation}
which satisfy the commutation relations $[\eta_x,\eta_y]=il^{*2}$ and $[R_x,R_y]=-il^{*2}$, while $[\eta_{x/y},R_{x/y}]=0$, in terms
of an effective CF magnetic length $l^\ast=\sqrt{2ps+1}l$.

As vortices are collective configuration of electrons, the newly introduced vortex operators in fact doubly count the physical degrees of freedom.
To heal this double counting, we need to restrict the dynamical variables to the physical sub-Hilbert space, which is subject to a
constraint $\chi(\mathbf q)|\textrm{phys}\rangle=0$, where $\chi(\mathbf q)=\sum_i \exp(-i\mathbf q\cdot \mathbf R_v)$ is the vortex density operator \cite{RevModPhys.75.1101}.
Reversing \eqref{eqrlcfe}, the electron density operator can be expressed as $\rho_e(\mathbf q)=\sum_i\exp[-i\mathbf q\cdot(\mathbf R_i+c\boldsymbol\eta_i)]$.

In the lowest LL, the Hamiltonian is simply given by the two-body interaction Eq. \eqref{eqprjH}. According to Eq. \eqref{eqrlcfe}, the magnetic field felt
by the CFs is reduced to $B/(2ps+1)$. Since the CF density is equal to the density of electrons, the CFs now fill completely $p$ CF LLs.
Therefore the huge degeneracy for fractional fillings is lifted and our calculation can be based on a state with these $p$ filled CF LLs. It serves
as the reference state for further diagrammatic approaches, such as the Hartree-Fock approximation, which we use to calculate the activation gap given by
\begin{equation}
\Delta=\langle \mathbf p+PH|H^p|\mathbf p+PH\rangle-\langle\mathbf p|H^p|\mathbf p\rangle,
\end{equation}
where $|\mathbf p\rangle$ stands for the ground state with $p$ filled LLs and $PH$ symbolizes a widely separated quasiparticle-quasihole pair.
In practice, the quasi-particle and quasi-hole are so far away that they have no correlations and both of their gaps (called as charged gaps) are computed
individually in the single-particle/hole excitation Hilbert space. The activation gap is the sum of two,
\begin{equation}
\Delta=\Delta_P+\Delta_H,
\end{equation}
where their explicit expressions are given by creating a particle in the $p$-th CF LL by applying $d^\dagger_p|\mathbf p\rangle$ or a hole in the $(p-1)$-th CF
LL via $d_{p-1}|\mathbf p\rangle$,
\begin{align}
\Delta_P&=\langle\mathbf p+P|H^p|\mathbf p+P\rangle=\langle \mathbf p|d_p H^pd_p^\dagger|\mathbf p\rangle,\\
\Delta_H&=\langle\mathbf p+H|H^p|\mathbf p+H\rangle=\langle \mathbf p|d^\dagger_{p-1} H^pd_{p-1}\mathbf p\rangle.
\end{align}
The above picture naturally follows from the CF transformation. However the Hartree-Fock ground state $|\mathbf p\rangle$ does not obey the physical
constraint $\chi(\mathbf q)|\mathbf p\rangle=0$. If we plug in the projected Hamiltonian Eq. \eqref{eqprjH} and the electron density $\rho_e$, this naive
procedure therefore suffers from strong corrections. A practical solution is to employ the preferred density $\rho^p=\rho_e-c^2\chi$ instead of $\rho_e$ in the Hamiltonian.
This density has the merits of complying with Kohn's theorem and giving the correct charge in the small-$q$ limit \cite{RevModPhys.75.1101}. The Hamiltonian is finally written as
\begin{equation}
H^p=\frac{1}{2}\sum_{\mathbf q}\rho^p(\mathbf q)V_{\textrm{eff}}(\mathbf q)\rho^p(-\mathbf q)\label{eqLLLH}
\end{equation}
and will serve as the starting point in our calculations presented in the following sections.

\section{A variational metric in the Hamiltonian theory}\label{scmrht}

\subsection{The deformed liquid state}

Quantum Hall systems with anisotropy are of particular interest due to possible phase transitions and competitions between them.  In particular, the deformation of the
Laughlin state is a good starting point for such systems. In this section, we show how Haldane's geometric point of view \cite{PhysRevLett.107.116801} can be implemented in the
Hamiltonian theory and provide interpretation of generalized pseudo-potentials in this language.

In an isotropic translationally invariant quantum Hall system, any two-body interaction $ \hat V$ can be expanded in terms
of the relative-angular-momentum basis, $ \hat V=\sum_{m} V_m|m\rangle\langle m|$, where $|m\rangle$ is the two-body state
with relative angular momentum $m$  \cite{PhysRevLett.51.605}. For fermions only odd values of $m$ are relevant to insure the antisymmetry of the wave function.
The Laughlin state for filling $\nu=1/q$ is characterized as the
unique zero-energy eigenstate of a certain class of Haldane's pseudo-potential \cite{PhysRevLett.51.605}, $H_M=\sum_{m=1}^{q-2} V_mP_m$, where $P_m$
is a projection operator $|m\rangle\langle m|$ to the space with relative angular momentum $m$ [see Eq. \eqref{eqgepp} for explicit forms] and pseudo-potentials $V_m$
are arbitrary positive energy coefficients. In an anisotropic but translationally invariant system, the interaction is no longer diagonal in the angular-momentum
basis, and a generalization of Haldane's pseudo-potentials has recently been proposed to incorporate those off-diagonal parts $|m+n\rangle\langle m|$
($n\ne 0$) \cite{PhysRevLett.118.146403}:
\begin{equation}
H_M=\sum_{m,n,\sigma}V^\sigma_{m,n}P^\sigma_{m,n}.\label{eqexpgepp}
\end{equation}
$\sigma=\pm$ is for symmetric and anti-symmetric combinations of $|m+n\rangle\langle m|$ and $|m\rangle\langle m+n|$ in order to make the interaction
Hermitian. $P^\sigma_{m,n}$ can be viewed as a complete basis for any two-body translationally invariant interaction $V_{\textrm{eff}}(\mathbf q)$ and
$V^\sigma_{m,n}$ are the expanding coefficients. The diagonal components $P^+_{m,0}=P_m$ are Haldane's original pseudo-potentials and the other components
are referred to as off-diagonal pseudo-potentials. In their presence, no model wave function is known as the zero-energy ground state. The off-diagonal parts
serve as perturbations to the liquid state. Their relative magnitude with respect to the diagonal pseudo-potentials, which define the Laughlin state,
provides good criterion for stability issues \cite{PhysRevB.96.195140,PhysRevB.97.035140}.

It is pointed out by Haldane \cite{PhysRevLett.107.116801} that the quantum Hall system may have a hidden variational geometric parameter, parameterized
by a metric. This can be manifested from the definition of Laughlin states according to pseudo-potentials. A metric $g$ can be introduced in its definition
and therefore, a family of generalized Laughlin wave functions is obtained,
\begin{equation}
P_m(g)|\Psi^q(g)\rangle=0, m<q.\label{eqdfanlau}
\end{equation}
The exact form of this generalized Laughlin state has been constructed \cite{PhysRevB.85.115308} by combining the metric with guiding center coordinates
to form ladder operators. For an isotropic system, the kinetic part and the interaction part of the Hamiltonian have the same flat metric. The effective
interaction is therefore isotropic. The Laughlin state $|\Psi(g)\rangle$ with a flat metric is then the most favored state. When the kinetic part
is perturbed by an anisotropic mass tensor or the interaction deviates from the isotropic case, the flat metric in the wave function will change to an
anisotropic one in order to minimize the energy. Two approaches have been established to find the optimal metric in this situation. For example, in the
presence of a parallel magnetic field, the optimal metric is obtained by finding the anisotropic Laughlin state $|\Psi(g)\rangle$ with highest overlap
with the numerical exact diagonalization result \cite{PhysRevB.87.245315}. Or it can be determined by finding the set of generalized pseudo-potentials
with minimal coefficients for the off-diagonal components of $P^\sigma_{m,n}(g)$ \cite{PhysRevB.96.195140}. As we have mentioned below Eq. \eqref{eqexpgepp},
any effective potential can be expanded in terms of $P^\sigma_{m,n}(g)$, by varying $g$, the off-diagonal components may have minimum expansion coefficients.

Now we introduce the metric variable into the Hamiltonian theory formalism. As introduced in the previous section,
the Hamiltonian theory is based on the fact that CFs fill $p$
LLs. For a charged particle under magnetic field, the cyclotron  and guiding-center coordinates satisfy the following commutation relation:
\begin{equation}
[\eta^a,\eta^b]=il^{\ast2}\epsilon^{ab},\quad [R^a,R^b]=-il^{\ast2}\epsilon^{ab}.
\end{equation}
They can be combined with a complex vector $v_\alpha$ to form ladder operators which define the CF LLs and their angular momenta,
\begin{equation}
\hat a=\frac{v_a\eta^a}{l^\ast},\, \hat a^\dagger=\frac{v^\ast_a\eta^a}{l^\ast},\quad \hat b=\frac{v^\ast_aR^a}{l^\ast},\, \hat b^\dagger=\frac{v_aR^a}{l^\ast},
\end{equation}
where the complex vector $v$ satisfies $v_av_b^\ast-v_a^\ast v_b=-i\epsilon_{ab}$. Here and in the remainder of this paper, we use Einstein's convention according to
which we sum over repeated indices of co- and contra-variant vectors.
A complete basis of the one-body Hilbert space is defined
as $|m,n\rangle\propto(\hat a^\dagger)^n(\hat b^\dagger)^m|0\rangle$. By changing the form of this vector, the eigenstates of ladder operators exhibit different orbital
shapes of $\boldsymbol \eta$ and $\mathbf R$. In general, the complex vector multiplied with the cyclotron coordinate and that with the guiding center
coordinate are different, as is shown in the original idea by Haladane of anisotropic Laughlin states. However, in the Hamiltonian theory case, both of the
two CF coordinates come from the guiding-center coordinates of electrons, because the pseudo-vortex is only an auxiliary variable composed of electronic degrees of freedom,
so that $\boldsymbol\eta$ and $\mathbf R$ are related to the same vector $v$. A metric $g$ conserving the area $\det(g)=1$
is associated with such a complex vector,
\begin{equation}
g_{ab}=v_av^\ast_b+v^\ast_av_b.
\end{equation}
When $v=(1,i)/2$, the metric is flat and the corresponding ladder operators are those defined by Shankar and Murthy \cite{RevModPhys.75.1101}.
As one varies the vector $v$, the metric associated with it is a direct reflection of the internal geometry.

The above construction can be viewed as building an anisotropic liquid phase from the Hamiltonian theory, parallel to the construction of anisotropic
Laughlin wave functions \cite{PhysRevB.85.115308}. There the metric is also parameterized by a complex vector. The polynomial factors are replaced by the
creation operators $\hat b^\dagger$ of electronic guiding centers
\begin{equation}
\Psi^q(g)=\prod_{i<j}(\hat b^\dagger_i-\hat b^\dagger_j)^q \Psi_{LLL},
\end{equation}
where $\Psi_{LLL}$ is the lowest Landau level wave function annihilated by $\hat a_i$ and $\hat b_i$ for every electron $i$, $\hat a_i\Psi_{LLL}=0$ and $\hat b_i\Psi_{LLL}=0$. This
definition is easily shown to be equivalent to Eq. \eqref{eqdfanlau}. Since the cyclotron and guiding-center coordinates are true degrees of
freedom, they are free to choose different metrics. The metric of the guiding center determines the geometry of the liquid state while the cyclotron metric
determines the LL shape. In contrast, in the Hamiltonian theory, the electronic guding-center metric induces that of the vortices.
There is therefore a common metric for the CF cyclotron and guiding-center coordinates.
The reference state of $p$ filled CF LLs, built using anisotropic
CF cyclotron coordinate, can thus be regarded as the operator description of the anisotropic Laughlin wave functions.

\subsection{The response of the metric to anisotropy}\label{scrlman}

Now we study how this metric responds to anisotropy in the effective interaction. First, in the Hamiltonian theory the ladder operators $\hat a$ and $\hat b$
determine the matrix entries
for the interaction $V_{\textrm{eff}}$. Therefore, the first point we need to verify is the Hartree-Fock nature of the Hamiltonian theory, namely whether
the anisotropy changes the structure of the Hartree-Fock ansatz. According to the CF construction, the ground state consists of $p$ filled LLs.
We need to know if the interaction $V_{\textrm{eff}}$ causes inter-Landau level transitions and thus mixes different Landau levels. In the isotropic case,
the answer is negative because, as we show below,
\begin{equation}
\langle\mathbf p|d_f H^pd_i^\dagger|\mathbf p\rangle\sim\delta_{fi},\ \textrm{for isotropic } V_{\textrm{eff}}(\mathbf q)=V(q),
\end{equation}
i.e. on the Hartree-Fock level the states $d_i^{\dagger}|\mathbf p\rangle$, obtained by adding CF to state in an empty level, are eigenstates of the Hamiltonian.
The state $d_i^\dagger|\mathbf p\rangle$ is obtained by adding a CF particle at an empty level $i$. It is also possible to show that the above equation holds
for creating a hole in filled levels. However, this relation strongly relies on the rotation invariance of the interaction $V$. When rotation symmetry is broken,
the interaction leads to mixing of LLs. And in the following we show that this mixing effect is proportional to the off-diagonal parts of the
generalized pseudo-potentials.

As long as the anisotropy is small, one can still assume that the ground state is composed of $p$ filled CF LLs because it is protected by the quasi-particle and quasi-hole gaps.
Hamiltonian \eqref{eqLLLH} is written in second quantization as
\begin{equation}
H^p=\frac{1}{2}\sum_{1,2,3,4}V_{\textrm{eff}}(\mathbf q)\rho^p_{12}(\mathbf q)\rho^p_{34}(-\mathbf q)d^\dagger_1d_2d^\dagger_3d_4.
\end{equation}
The subscripts $1,2,3,4$ are abbreviations of $(m,n)$ corresponding to the CF states defined before. $d_i$ and $d^\dagger_i$ are annihilation and
creation operators of CFs in this basis, and $\rho^p_{ij}(\mathbf q)$ is the matrix element of the preferred density operator:
\begin{equation}
\rho^p(\mathbf q)_{m,n;m',n'}=\langle m,n|\rho_e(\mathbf q)-c^2\chi(\mathbf q)|m',n'\rangle.
\end{equation}
The above matrix has a product structure for the indices of angular momenta and Landau levels,
$\rho^p(\mathbf q)_{m,n;m',n'}=\rho^p(\mathbf q)_{m,m'}\otimes\rho^p(\mathbf q)_{n,n'}$. With the Hartree-Fock assumption, the activation gap can be calculated
using Wick's theorem. The matrix element of the Hamiltonian between two one-particle excitation states is:
\begin{align}
\langle p|d_f H^pd_i^\dagger|p\rangle=&\frac{1}{2}\int \frac{d^2q}{4\pi^2} V_{\textrm{eff}}(\mathbf q)\rho^p_{12}(\mathbf q)\rho^p_{34}(-\mathbf q)\nonumber\\
&\times \sum_{1,2,3,4}[\delta_{f1}(1-n_1)\delta_{23}(1-n_2)\delta_{4i}(1-n_4)\nonumber\\&
\qquad -\delta_{f3}(1-n_3)\delta_{14}n_4\delta_{2i}(1-n_2)],
\end{align}
 and a similar expression for hole excitations. Because of the product structure of $\rho$, it can be shown that the interaction preserves the CF angular momentum.
 Using the identity $\exp(-i\mathbf q\cdot\mathbf R)\times\exp(i\mathbf q\cdot\mathbf R)=\mathbf 1$, the above expression is thus proportional to $\delta_{m_f,m_f}$.
 It only mixes different Landau levels of the composite fermion. By summing up all guiding center parts, one obtains:
\begin{align}
\langle p|d_f H^pd_i^\dagger|p\rangle=&\frac{1}{2}\int \frac{d^2q}{4\pi^2} V_{\textrm{eff}}(\mathbf q)\Theta(n_f-p)\Theta(n_i-p)\nonumber\\
&\left[\sum_{n_2=p}^\infty\rho^p_{n_fn_2}(\mathbf q)\rho^p_{n_2n_i}(-\mathbf q)\right.\nonumber\\
&\qquad-\sum_{n_1=0}^{p-1} \rho^p_{n_1n_i}(\mathbf q)\rho^p_{n_fn_1}(-\mathbf q)\Bigg]
\delta_{m_f,m_i},\label{eqmxeleo}
\end{align}
and a similar formula for hole excitations. $\Theta$ is the step function with $\Theta(x<0)=0$ and $\Theta(x\geq 0)=1$. In order
to study the role of anisotropy in $V_{\textrm{eff}}$, we need explicit expressions for the matrix elements of the density operator $\rho_{n1,n2}$. This can be
achieved with the help of coherent states.

From the basis of one-particle states,
\begin{equation}
|mn\rangle=\frac{(\hat b^\dagger)^m}{\sqrt{m!}}\frac{(\hat a^\dagger)^n}{\sqrt{ n!}}|00\rangle,
\end{equation}
one can construct coherent states that are eigenstates of the ladder operators. Taking e.g. the operators $\hat a$, the coherent states are defined as:
\begin{equation}
|z\rangle=e^{a^\dagger z}|0\rangle=\sum_{n=0}^{\infty}\frac{|n\rangle}{\sqrt{n!}}z^n,\quad \langle \bar z|z\rangle=e^{z\bar z}.
\end{equation}
Here the normalization is not fixed to one in order to compare the coefficients with the matrix elements later. In the definition of coherent state,
the guiding-center coordinates naturally appear in the exponent. Thus the expectation value of the density operator for coherent states can be expressed as:
\begin{equation}
\langle \bar z|e^{-i\mathbf{q\cdot}\boldsymbol\eta}|z\rangle=\sum_{n_1=0}^{\infty}\sum_{n_2=0}^\infty
\frac{\bar z^{n_2}}{\sqrt{n_2!}}\frac{z^{n_1}}{\sqrt{n_1!}}\langle n_2|e^{-i\mathbf{q\cdot}\boldsymbol\eta}|n_1\rangle.\label{eqcohexp}
\end{equation}
On the other hand, the above expectation can be obtained from expanding the density operator in terms of ladder operators:
\begin{equation}
q_a\eta^a=q_a\delta^a_b\eta^b=q_a(v^av_b^\ast+v^{a\ast}v_b)\eta^b=q_av^a \hat a^\dagger+q_av^{a\ast}\hat a,
\end{equation}
where the contra-variant vector $v^a=g^{ab}v_b$ is defined with the help of the inverse of the metric, $g^{ab}=(g_{ab}^{-1})$. In the second equality we
use the property $g^{ac}g_{cb}=\delta^a_b\to v^av_b^\ast+v^{a\ast}v_b=\delta^a_b$. So the expectation value can be written as:
\begin{align}
\langle \bar z|e^{-i\mathbf{q\cdot}\boldsymbol\eta}|z\rangle&=\langle \bar z|\exp\left[-il^\ast(q_+a^\dagger+q_-a)\right]|z\rangle\nonumber\\
&=\langle \bar z-il^\ast q_+|z-il^\ast q_-\rangle e^{q^2l^{\ast2}/4}\nonumber\\
&=\exp\left[\bar zz-il^\ast(q_+z+q_-\bar z)\right]e^{-q^2l^{\ast2}/4},
\end{align}
where $q_+=q_av^a,q_-=q_av^{a\ast}$. Comparing the the above equation with Eq. \eqref{eqcohexp}, one obtains:
\begin{equation}
\langle n_2|e^{-i\mathbf{q\cdot}\boldsymbol\eta}|n_1\rangle=\sqrt{\frac{n_2!}{n_1!}} e^{-x/2}\left(\frac{-iq_+l^\ast}{\sqrt 2}\right)^{n_1-n_2}L^{n_1-n_2}_{n_2}(x),
\end{equation}
where $x=q_aq_bg^{ab}l^\ast/2$ and $L^\alpha_n$ are associated Laguerre polynomials.
The result is valid for $n_1\ge n_2$. For $n_1\le n_2$, the matrix elements are obtained from the complex
conjugation $\langle n_2|e^{-i\mathbf{q\cdot}\boldsymbol\eta}|n_1\rangle=\langle n_1|e^{i\mathbf{q\cdot}\boldsymbol\eta}|n_2\rangle$.

With these expressions for $\rho^p(\mathbf q)$ at hand, we are able to evaluate the Hamiltonian in one-particle/hole excitation space.
The interaction preserves particle numbers. Therefore for the excitation of a CF, both the initial state and the final state should
have the quasiparticle on an empty CF LL, such that there is no mixing between hole excitations and particle excitations. Now if we choose the metric
$g_{ab}$ to be flat, i.e. $v=(1,i)/\sqrt 2$, the reduced matrix behaves like $\rho^p(\mathbf q)_{n1,n2}\sim \exp[-i(n_1-n_2)\phi]$, where $\phi$ is the
angle between $\mathbf q$ and the $x$-axis. Hence according to Eq. \eqref{eqmxeleo} we observe that
\begin{equation}
\langle p|d_f H^pd_i^\dagger|p\rangle\propto\int d\phi V_{\textrm{eff}}(\mathbf q)e^{-i(n_f-n_i)\phi}.
\end{equation}
This actually gives an angular decomposition of the effective potential, which means, if $V_{\textrm{eff}}$ is isotropic, there is no mixing between
different single-particle excitation states, as mentioned above. The Hartree-Fock assumption is stable and self-consistent. However, if $V_{\textrm{eff}}$ is anisotropic,
it will induce transition from one CF LL to another. When the transition is small, the anisotropy is simply a perturbation to the picture of $p$ filled CF LLs.
When the transition is sufficiently large, the spacing between Landau levels may even close, in which case the liquid state is no longer stable.

In particular, we can also interpret this in the language of generalized pseudo-potentials. The explicit expressions of them are as follows \cite{PhysRevLett.118.146403}:
\begin{align}
P^+_{m,n}(g)&=\lambda_n\mathcal N_{m,n}\left(L^n_m(|q|^2)e^{-\frac{1}{2}|q|^2}q_+^n+\textrm{c.c}\right),\\
P^-_{m,n}(g)&=-i\mathcal N_{m,n}\left(L^n_m(|q|^2)e^{-\frac{1}{2}|q|^2}q_+^n-\textrm{c.c}\right),\label{eqgepp}
\end{align}
where $\lambda_n=1/\sqrt 2$ when $n=0$ and $\lambda_n=1$ for $n\ne 0$. $\mathcal N_{m,n}$ is normalization for this expansion of two-body interaction.
They are chosen to be real and the components $P^-_{m,n}$ with negative superscripts are only non-vanishing for $n\ne 0$ and therefore do not exist in
an isotropic case. Further we have $|q|^2=q_aq_bg^{ab}$ and $q_+$ is defined as before. When choosing the flat metric $g=\mathbf 1$, $q_+=|q|e^{i\phi}/\sqrt 2$,
one can immediately observe that a generalized pseudo-potential $P^\pm_{m,n}$ introduces transition from a CF LL $n'$ to $n'\pm n$. Notice, however, that since the
CF does not have the same magnetic length as an electron, the off-diagonal pseudo-potential $P^\pm_{m,n}$ are not one-to-one correspondence
to different LL transitions, but they rather mix all the CF LLs separated by a multiple of $n$. In Ref. \cite{PhysRevB.96.195140},
the optimal $g$ is chosen such as to minimize the off-diagonal components $n\ne 0$, because they are perturbations to the Haldane's pseudo-potentials which
define the liquid state. The Hamiltonian theory provides us a complementary picture: the
off-diagonal pseudo-potentials cause transitions between CF LLs and thus destroy the Hartree-Fock nature of the ground state. However, the variation of the metric $g$ allows
us to minimize CF LL mixing and thus to save the picture of $p$ filled CF LLs. Since this transition amplitude is proportional to the off-diagonal pseudo-potentials,
the optimal metric obtained in this way agrees with that of Ref. \cite{PhysRevB.96.195140}, as we show in the following section.

\begin{figure}
  \centering
  \includegraphics[width=0.35\textwidth]{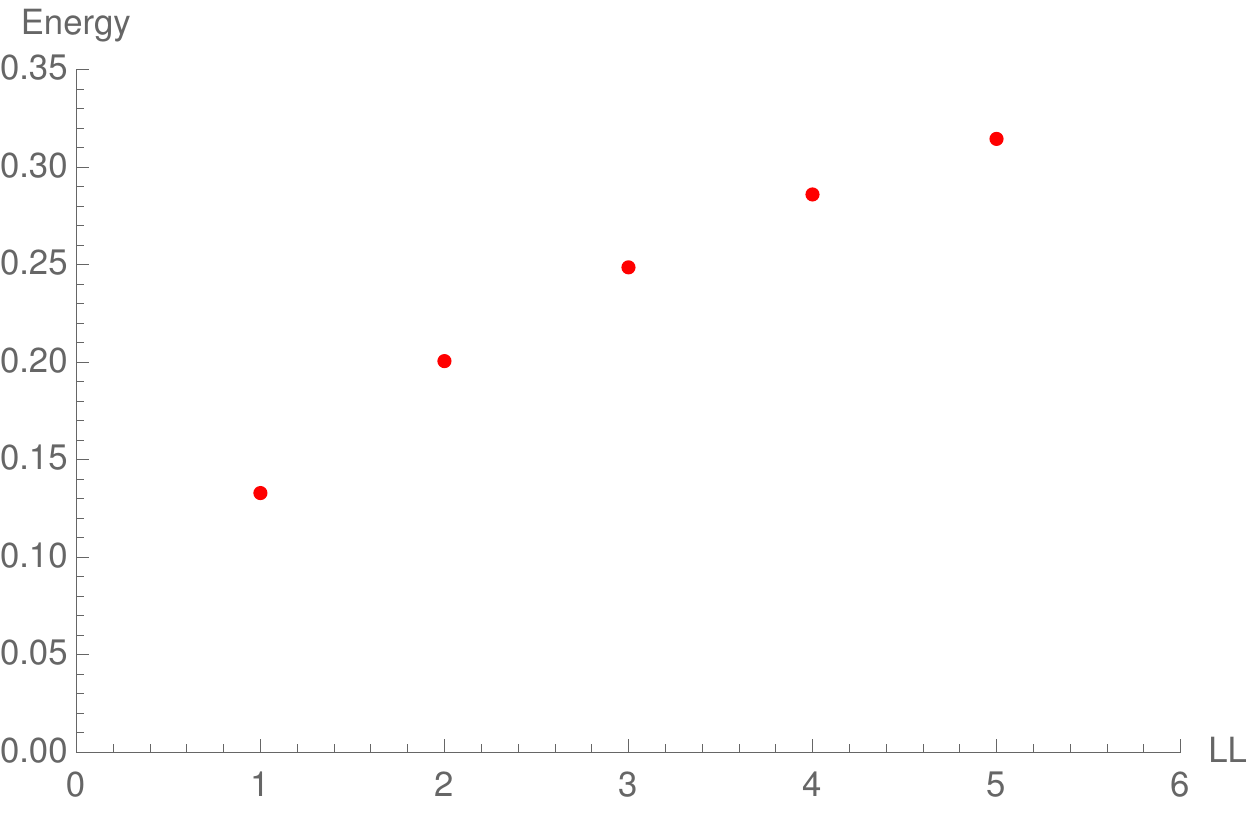}\\
  \caption{The energy of creating a composite fermion with respect to the CF LL index for $\nu=1/3, \tan\theta=0,\Omega/\omega_c=2$, in the unit of $e^2/\varepsilon l$ with $\varepsilon$ the dielectric constant.
  }\label{figenerlin}
\end{figure}

Making use of the operator expression of $\rho(\mathbf q)\rho(-\mathbf q)$, Eq. \eqref{eqmxeleo} is further simplified to
\begin{align}
\langle p|d_f H^pd_i^\dagger|p\rangle=&\frac{1}{2}\int \frac{d^2q}{4\pi^2} V_{\textrm{eff}}(\mathbf q)\bigg\{(1+c^4f^2)\delta_{n_f,n_i}\nonumber\\
&-c^2f\cos\left[\mathbf{q\cdot}\boldsymbol\eta\left(c-\frac{1}{c}\right)
\right]_{n_fn_i}\\
&-2\sum_{n_1=0}^{p-1}
\rho^p_{n_1n_i}(\mathbf q)\rho^p_{n_fn_1}(-\mathbf q)\bigg\}\delta_{m_f,m_i},\nonumber
\end{align}
which allows us to calculate the matrix elements of $H^p$. Here, $f= \exp(-q^2l^2/4c^2)/\exp(-q^2l^2/4)$ is the vortex form factor,
which takes into account the difference between the electric and the vortex charge.
The Hilbert space is infinite and we have an infinite number of CF LLs,
which makes it impossible to diagonalize the single-particle excitations. In order to obtain a quantitative result, we truncate the number of CF LLs.
As pointed out in Ref. \cite{PhysRevB.60.13702}, even if the preferred density is assumed to take into account non-perturbative corrections,
the physical constraint is still not imposed faithfully, and higher CF Landau levels are actually nonphysical. A reasonable approximation is to examine up
to which CF LL the following relation holds:
\begin{equation}
\langle\mathbf p+m|H^p|\mathbf p+m\rangle-\langle\mathbf p|H^p|\mathbf p\rangle\approx |m-p|\hbar\omega_\ast+\textrm{cst.},
\end{equation}
where $|\mathbf p+m\rangle$ means the state obtained by creating a quasi-particle ($m\ge p$) or a quasi-hole ($m<p$) of LL $m$
on top of the $p$-filled ground state $|\mathbf p\rangle$. The above formula derives from the fact that the CF experiences a reduced magnetic field.
For high indices of the CF LL, this relation is no longer valid (see Fig. \ref{figenerlin}). A reasonable criterion is to take into account only $2p+1$
CF Landau levels for a filling $\nu=p/(2ps+1)$. For Landau levels higher than $2p+1$, their energies become closer and closer and approach a continuum limit.

\section{Activation gap in tilted magnetic field}\label{scaps}

\begin{figure}
  \centering
  \includegraphics[width=0.45\textwidth]{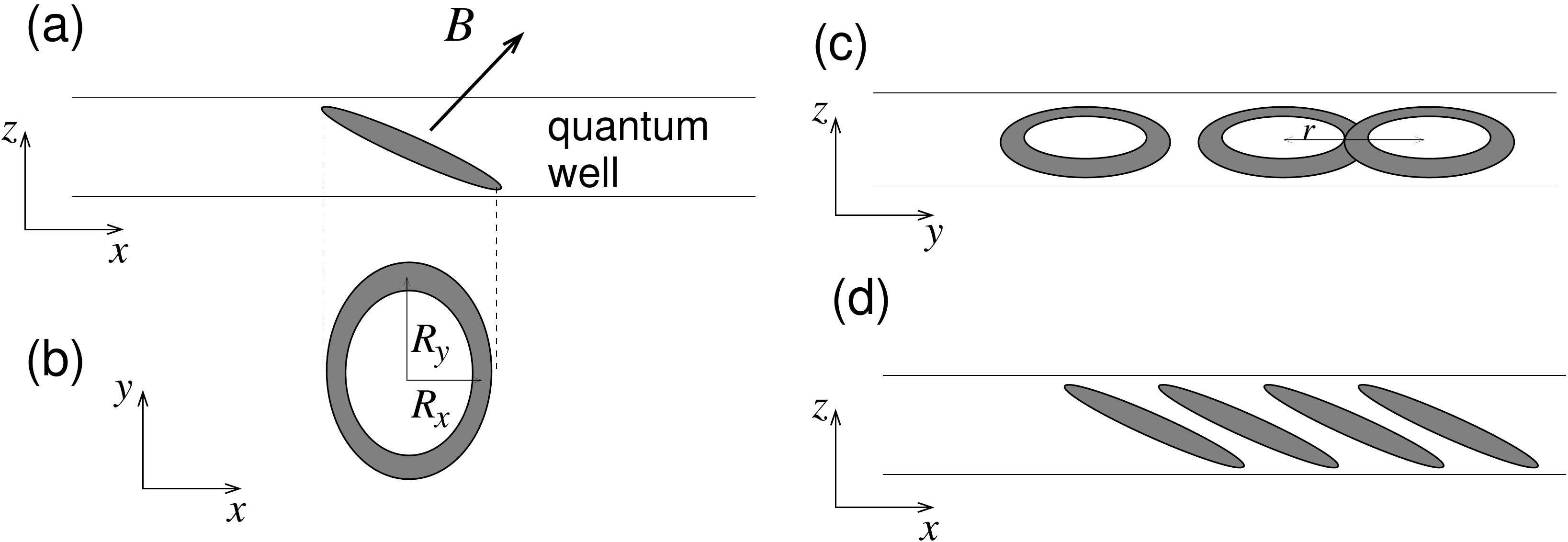}\\
  \caption{ Effect of a tilted magnetic field in a wide quantum well. (a) The wave functions tend to be in a plane roughly perpendicular to the total magnetic field. In contrast to
  a perfect 2D system, the wide quantum well allows for the tilt of the plane of cyclotron motion. (b) Resulting cyclotron motion projected to the $xy$ plane. (c) Adjacent wave functions
  of electrons in the $y$-direction. (d) Adjacent wave functions
  of electrons in the $x$-direction -- their wave-function overlap is largely reduced as compared to that in the $y$-direction [panel (c)].
  }\label{fig:widewell}
\end{figure}

For an ideally 2D system, adding an in-plane magnetic field has no effect, apart from an enhancement of the global Zeeman effect,
since it only causes movements in the restricted perpendicular direction.
However, in realistic systems, usually the sample has a finite thickness. While the motion of the electrons remains restricted in the $z$-direction,
they can respond to the parallel component of the magnetic field. Indeed, the electrons tends to perform its cyclotron motion in a plane perpendicular to the total
magnetic field such that the plane of motion is tilted, as sketched in Fig. \ref{fig:widewell}(a). The projection of the associated wave function is therefore an
ellipse, as shown in Fig. \ref{fig:widewell}(b). One thus notices already one origin of an anisotropy in the effective interaction potential. For a more quantitative
analysis, we model the restriction in the $z$-direction by a parabolic confining potential $m\Omega z^2/2$, which, together with the
kinetic energy in the perpendicular direction, needs to be added to the one-particle Hamiltonian,
\begin{equation}
H_z=\frac{\Pi_z^2}{2m}+\frac{m\Omega^2 z^2}{2}.
\end{equation}
Such a Hamiltonian is a quadratic form and can be solved exactly as a harmonic oscillator. To obtain the effective 2D motion, we project
the system to the ground state of this harmonic oscillator. The projection changes the interaction between electrons in a similar way of projecting the physics
to the lowest Landau level. In the absence of the in-plane magnetic field, the effective two-dimensional interaction has a cut-off at short distance:
\begin{equation}
V_\parallel (q)=\frac{2\pi e^2}{q}e^{(qlb)^2}\textrm{Erfc}(qlb),
\end{equation}
where $b=\omega_c/2\Omega$, $\omega_c=eB/m$ is the cyclotron frequency, and $bl$ is the typical length of this potential.

When the magnetic field is tilted towards the plane by an angle $\theta$, the cyclotron motion of the electron is also tilted towards the direction of the magnetic field, as already
mentioned above. In addition to the elliptic shape, also the relevant overlap between the wave functions is drastically altered -- while one maintains a substantial overlap between
the wave functions of adjacent electrons in the $y$ direction (perpendicular to the inplane component of the magnetic field), the tilted wave functions can be ``stacked'' in the
$x$-direction and their overlap substantially reduced [see Fig. \ref{fig:widewell}(c) and (d)].
As a consequence, the effective interaction is anisotropic with a space inversion symmetry. The effective
interaction for the $n$-th Landau level of two-dimensional electrons under a tilted magnetic field can be written as (see Ref. \cite{PhysRevLett.84.1288,PhysRevB.87.245315}
and the appendix \ref{aptm}):
\begin{align}
V_{\textrm{eff}}(\mathbf q)=&\int \frac{dq_z}{2\pi}\frac{4\pi e^2}{q^2}e^{-\frac{1}{2}\left[\frac{l_B^2 q_y^2\sin^2\tilde\theta}{l_+^2}+l^2_+(q_x\sin\tilde\theta-q_z\cos\tilde\theta)^2\right]}\nonumber\\
&\times e^{-\frac{1}{2}\left[\frac{l_B^2 q_y^2\cos^2\tilde\theta}{l_-^2}+l^2_-(q_x\cos\tilde\theta+q_z\sin\tilde\theta)^2\right]}\times\nonumber\\
&L^2_n\left[\frac{1}{2}\left(\frac{l_B^2 q_y^2\cos^2\tilde\theta}{l_-^2}+l^2_-(q_x\cos\tilde\theta+q_z\sin\tilde\theta)^2\right)\right],
\end{align}
where $l_+$ and $l_-$ are the typical lengths from the confining potential and the magnetic field respectively. $\tilde\theta$ is a function of the tilted angle $\theta$,
vanishing when there is no tilting \cite{PhysRevLett.84.1288}
\begin{equation}
\tan 2\tilde\theta=\frac{\tan 2\theta}{1-\tan^2\theta-\frac{\Omega^2}{\omega_c^2}}.
\end{equation}
The above interaction is invariant under $q_x\to -q_x,q_y\to -q_y$. From the decomposition of generalized pseudo-potentials, it has thus no component of
$P^{\pm}_{r,s}$ with $s$ odd since the latter does not have an inversion symmetry. According to the discussion in last section, this anisotropic interaction
only mixes CF LLs different by an even index.
\begin{figure}
\centering
\begin{subfigure}[b]{0.45\textwidth}
   \includegraphics[width=1\linewidth]{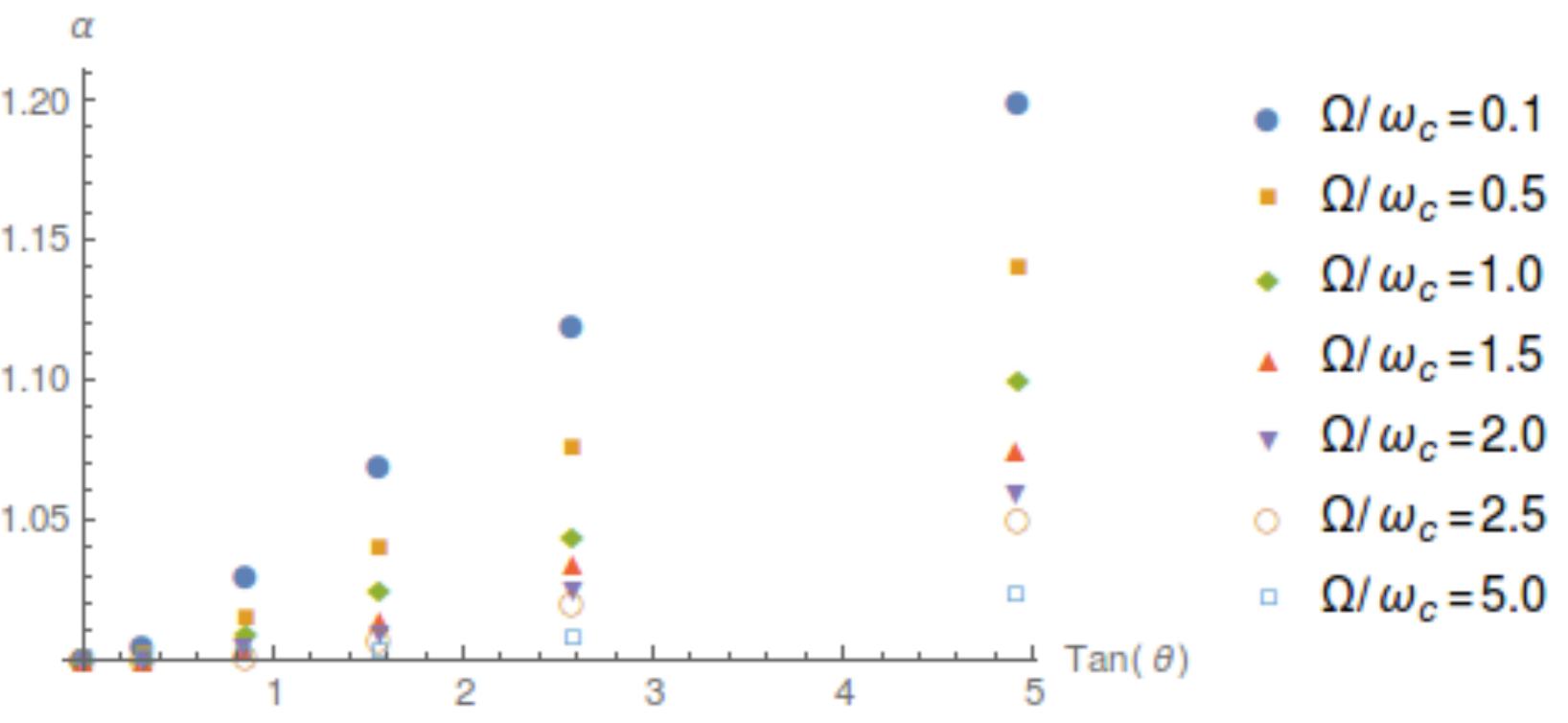}
   \caption{}
   \label{figme1o3}
\end{subfigure}

\begin{subfigure}[b]{0.45\textwidth}
   \includegraphics[width=1\linewidth]{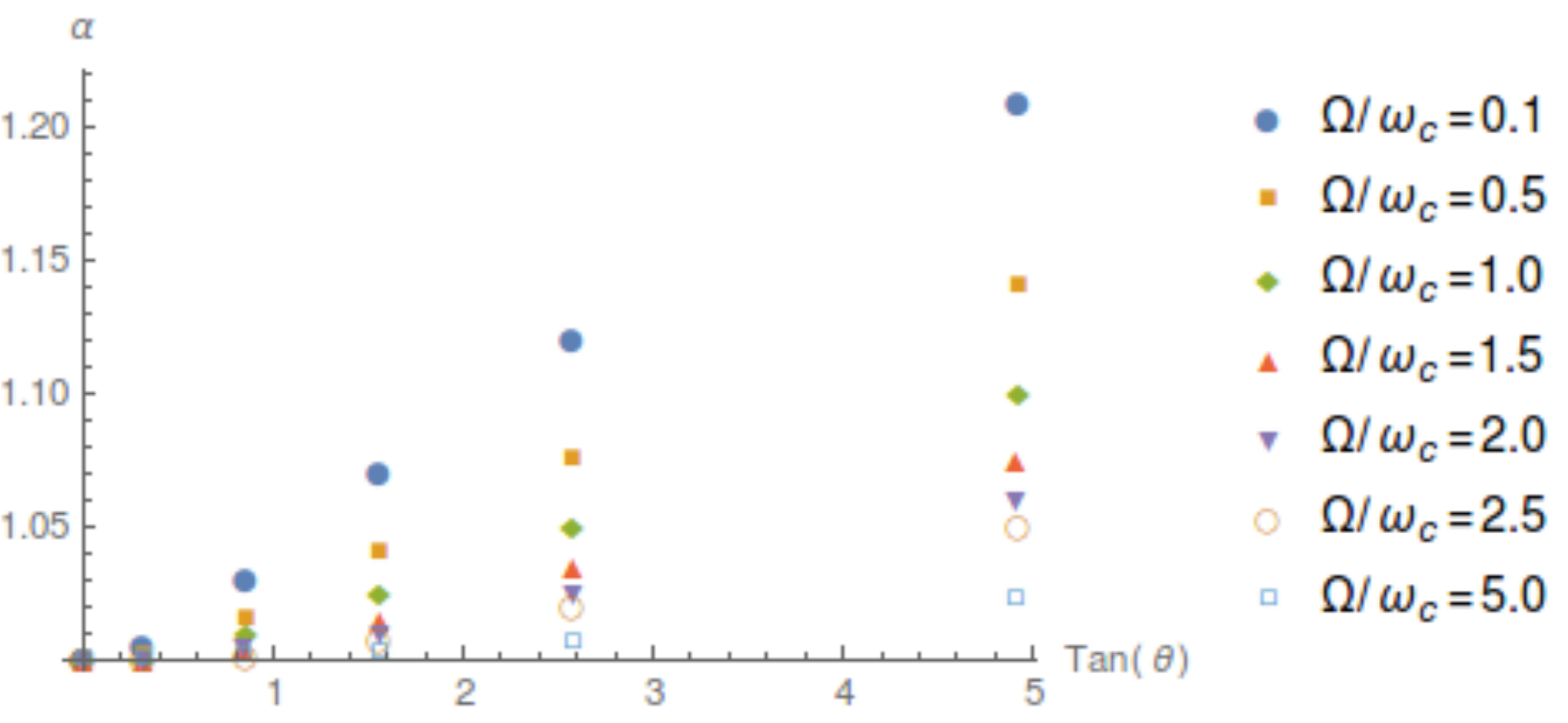}
   \caption{}
   \label{figme2o5}
\end{subfigure}

\caption{(a) Values of $\alpha$ of optimal metrics for $\nu=1/3$, at different confining potentials. (b) Values of $\alpha$ of optimal metrics for $\nu=2/5$.}
\end{figure}

We calculate the activation gaps and optimal metric for filling $\nu=1/3$. As mentioned in the last section, the anisotropy introduces transitions
between different Landau levels. The single-particle excitation matrix is defined as the expectation value of $V_{\textrm{eff}}(\mathbf q)$ when creating
a hole in a filled CF Landau level or a particle in an empty level
\begin{equation}
M^{\textrm{single}}_{m,n}=\langle\mathbf p+m|V_{\textrm{eff}}|\mathbf p+n\rangle.
\end{equation}
The spectrum of exciting a single particle/hole is obtained by diagonalizing this matrix. First, we calculate matrix elements in the flat metric case.
For $\nu=1/3$, the ground state has the $n=0$ Landau level filled. Holes are created in the $n=0$ Landau levels and particles are created in $n>0$ Landau levels.
The hole excitation has no mixing with the particle excitation because of particle number conservation. So the perturbation to the activation gap comes
from the transition from $n=1$ to higher CF Landau levels. According to the previous arguments, the first three states are deemed to be physical.
As the system has an inversion symmetry, the lowest Landau level that perturbs the activation gap is the $n=3$ level, at the brim of physical states.
Therefore, to obtain the leading perturbation, it is reasonable to take only the transition from $n=1$ to $n=3$ into account. The relevant Landau level
mixing amplitude is $M_{1,3}=M_{3,1}^\ast$. It is the direct reflection of the anisotropic effects.
\begin{figure}
\centering
\begin{subfigure}[b]{0.45\textwidth}
   \includegraphics[width=1\linewidth]{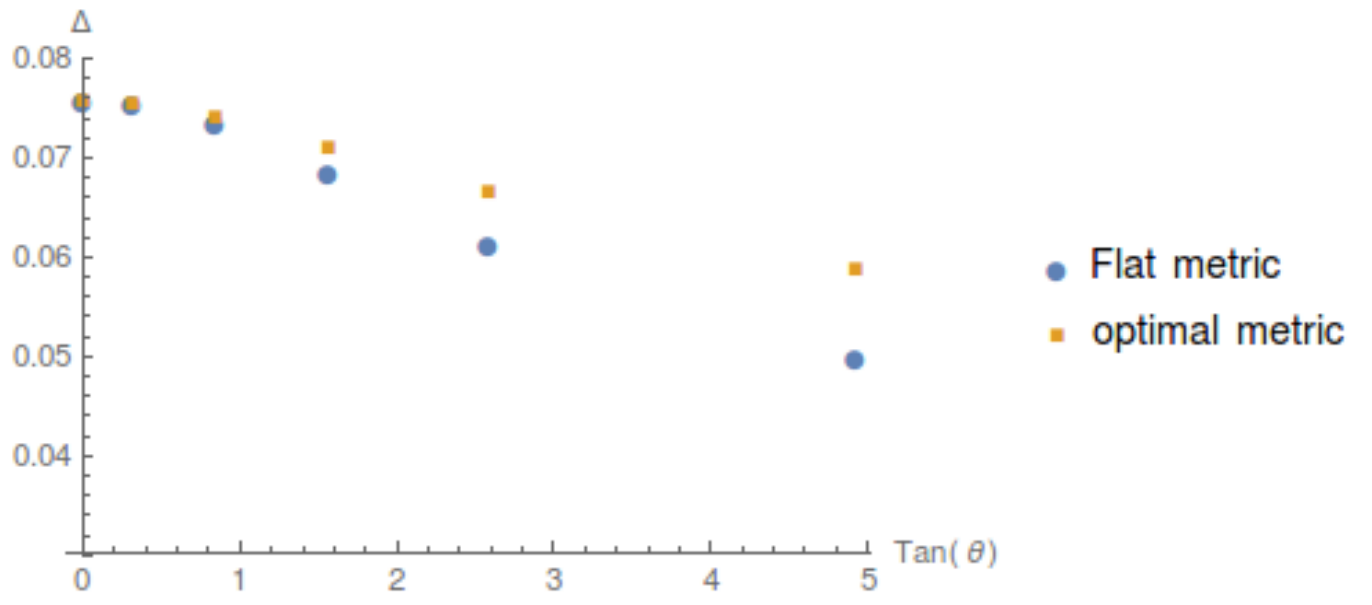}
   \caption{}
   \label{figact1o3op0d5}
\end{subfigure}

\begin{subfigure}[b]{0.45\textwidth}
   \includegraphics[width=1\linewidth]{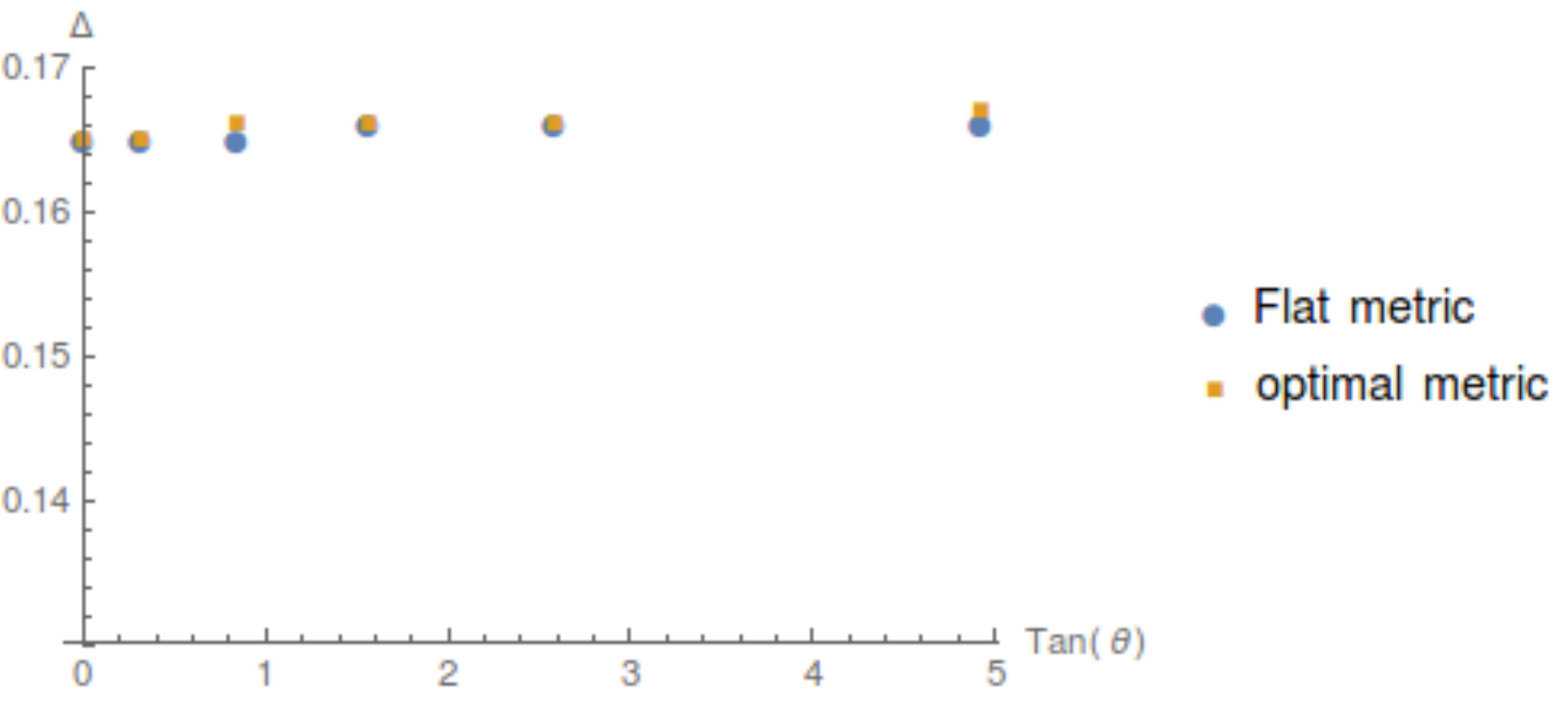}
   \caption{}
   \label{figact1o3op5}
\end{subfigure}

\begin{subfigure}[b]{0.45\textwidth}
   \includegraphics[width=1\linewidth]{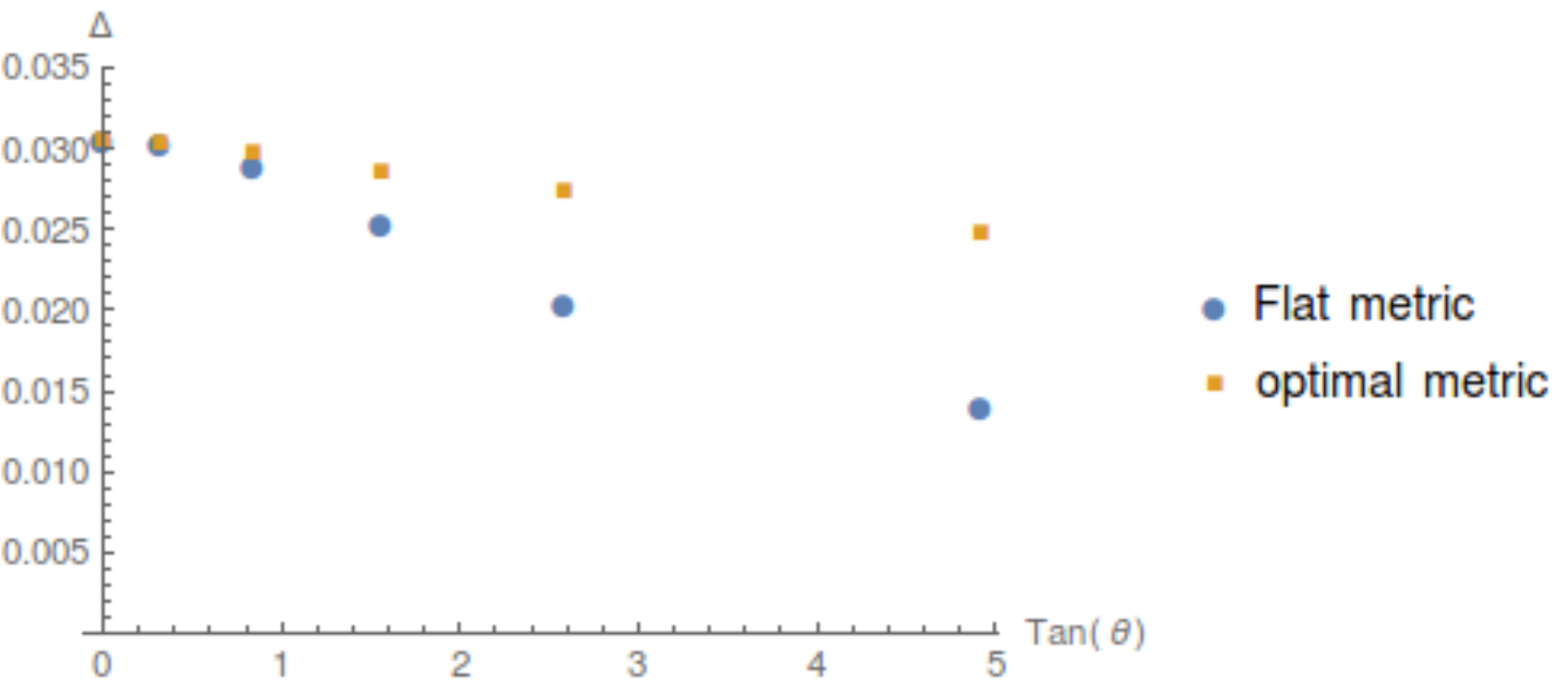}
   \caption{}
   \label{figact2o5op0d5}
\end{subfigure}

\begin{subfigure}[b]{0.45\textwidth}
   \includegraphics[width=1\linewidth]{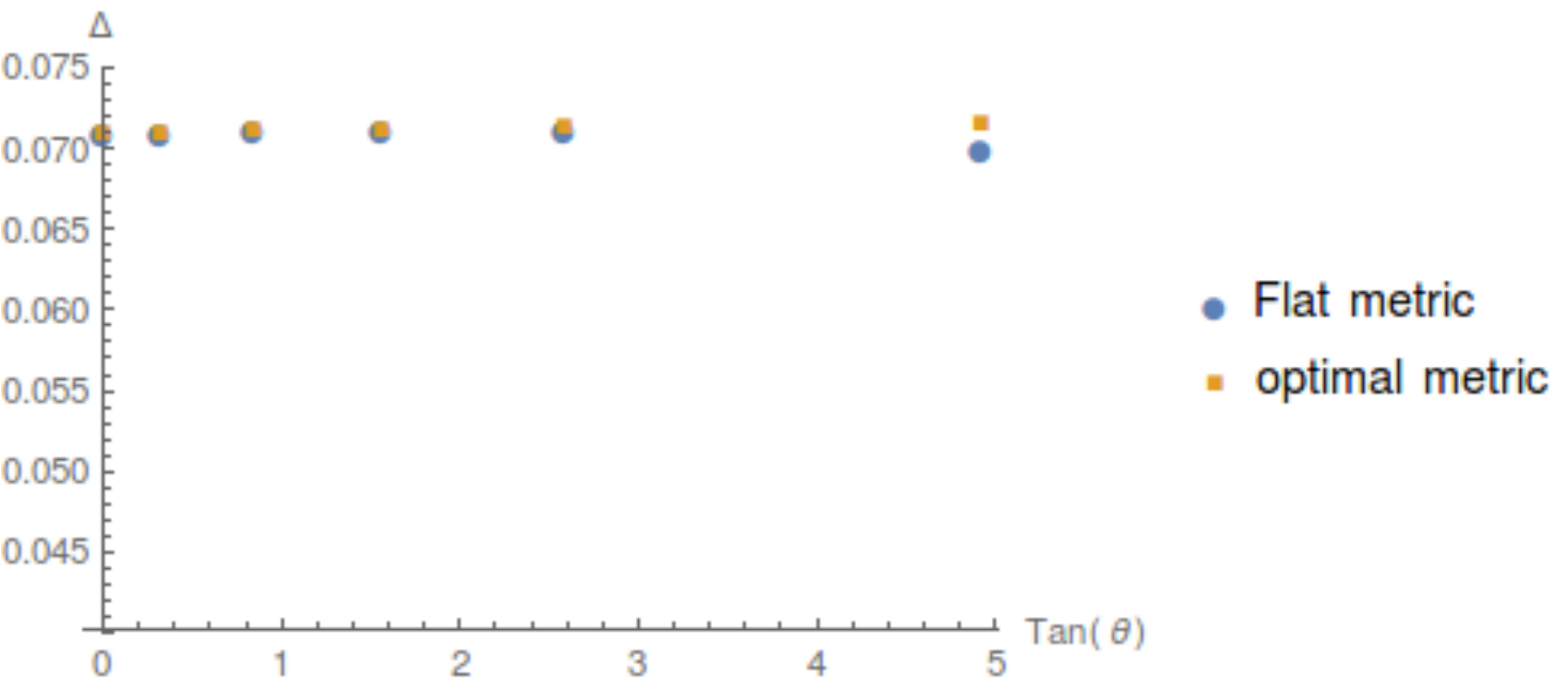}
   \caption{}
   \label{figact2o5op5}
\end{subfigure}

\caption{(a) The activation gap for $\nu=1/3$ of a weak confining potential $\Omega/\omega_c=0.1$. (b) The activation gap for $\nu=1/3$ of a strong confining
potential $\Omega/\omega_c=5$. (c) The activation gap for $\nu=2/5$ of a weak confining potential $\Omega/\omega_c=0.1$. (d) The activation gap for $\nu=2/5$
of a strong confining potential $\Omega/\omega_c=5$.}
\end{figure}

Let us now add the variational metric to the Hamiltonian theory. The optimal metric is the one which minimizes the inter-Landau level transition amplitude $M_{1,3}$.
As the in-plane magnetic field picks a preferred direction and the interaction is inversion invariant, the system has two principal axes. Due to this symmetry,
the ansatz of the metric takes the form:
\begin{equation}
g_{ab}=
\left(\begin{array}{ccc}
\alpha^2&0\\
0&1/\alpha^2
\end{array}\right).
\end{equation}
The corresponding complex vector of this metric can be parameterized by $v_a=(\alpha,i/\alpha)/\sqrt 2$. The results of $\alpha$ are shown in Fig. \ref{figme1o3}.
The optimal metric starts to deviate from one at around $\tan\theta=0.5$ ($\theta=27^\circ$) for all thicknesses. As the anisotropic effects come from the
movements in $z$ direction, we expect the metric is more prominent for large thickness. This is indeed the behavior in our results. The thicker the sample is, the more
rapidly the optimal metric, which is approximately linear in $\tan\theta$, grows with respect to the tilting angle. Notice, however, that the deformation
is always a small one.  To connect our calculation to tilted magnetic field experiments, for a wide quantum well of width $50$ nm \cite{PhysRevLett.108.196805},
the ratio between the confining frequency and the cyclotron frequency is estimated to be $\Omega/\omega_c=2/B$[T]. At the filling $\nu=1/3$ and $2/5$, the confining
frequency can be much smaller than the cyclotron frequency for such wide quantum wells, $\Omega/\omega_c\simeq 0.1\sim0.2$. So we set $\Omega/\omega_c=0.1$ as our
lowest value in calculation. We see that even at the extreme case, with $\Omega/\omega_c=0.1$ and $\tan\theta=5$, the optimal metric deviates from the flat one by only $20\%$.
Therefore, the deformation remains small. This agrees with the results
of generalized pseudo-potential analysis \cite{PhysRevB.96.195140}.

Figure  \ref{figact1o3op0d5} and \ref{figact1o3op5} show our results for the activation gap, for two different values of the confinement potential, as a function
of the tilt angle. The yellow squares indicate those calculated for the optimal metric evaluated for each value of the tilt angle, in comparison with the activation
gaps obtained from a purely flat metric (blue circles).  We note that for moderate parameters, the inter-Landau level transition $M_{1,3}$ is always much smaller
than the Landau level spacing $M_{1,1}+M_{0,0}$. For $\Omega/\omega_c=2$ and $\theta=35^\circ$, $M_{1,3}=0.002$ while $M_{1,1}+M_{0,0}=0.147$, the former is only one percent
of the latter. Notice that even in the extreme case of a wide quantum well $\Omega/\omega_c=0.1$ with $\tan\theta=5$, i.e. an inplane component of the magnetic field that is
five times larger than its perpendicular component, the ratio between $M_{1,1}+M_{0,0}=0.050$ and $M_{1,3}=0.017$ is roughly $1/3$ only, such that our variational metric
can still cope with this situation. Hence the activation is robust against the parallel magnetic field -- when the CFs
pick the optimal metric, the activation gap decreases much more slowly than that obtained from a flat metric even for weak confining potential and large tilting angles.

We also perform the same calculation for the $\nu=2/5$ filling [see Figs. \ref{figact2o5op0d5} and \ref{figact2o5op5}].
Among the physically relevant CF LLs, the lowest one perturbing the activation gap is the $n=4$ level.
The optimal metric, which we have calculated for $\Omega/\omega_c$ between $0.1$ and $5$, is almost the same as $\nu=1/3$ [see Fig. \ref{figme2o5}].
This is in accordance with our discussion in Sec. \ref{scrlman}. Indeed, all we are minimizing is the $\delta n=2$ transition, which is proportional
to the $P^\pm_{m,2}$ pseudo-potential components. When the metric is chosen to minimize these components, the transition amplitude is minimal, so that the
optimal metric for $\nu=1/3$ and $\nu=2/5$ should be close. On the other hand, as the activation gap itself for $\nu=2/5$ is smaller, the perturbation
due to the parallel magnetic field is stronger. One can see that for large tilting, where the LL mixing amplitude $M_{2,4}$ is of the same magnitude as
the Landau level spacing $M_{2,2}+M_{1,1}$, meaning that the Hartree-Fock ansatz receives strong correction.

\begin{table}
  \centering
  \caption{Matrix elements of $M^{\textrm{single}}$, the expectation value of $H^P$ in the single-particle/hole excitation space, for filling $\nu=2/5$
  at $\tan\theta=5,\Omega/\omega_c=0.1$, the most extreme case in our calculation.}
  \begin{tabular}{llllll}
  \hline
  \hline
  $M^{\textrm{single}}_{2,2}$  & $M^{\textrm{single}}_{4,4}$ & $M^{\textrm{single}}_{6,6}$
  & $M^{\textrm{single}}_{2,4}$ & $M^{\textrm{single}}_{2,6}$ & $M^{\textrm{single}}_{4,6}$\\
  \hline
  0.039 & 0.060 & 0.075 & -0.018& -0.002& -0.021\\
  \hline
  \hline
  \end{tabular}
  \label{tbmeM}
\end{table}

As a supplement to the truncation that we use, we also verify the perturbations from higher LLs. Take $\nu=2/5$ for example, the mixing between
the second and the sixth LLs $M_{2,6}$ is always one magnitude smaller than the mixing between the second and the fourth LLs $M_{2,4}$. In table \ref{tbmeM},
the extreme case of large tilting and thickness is listed. The matrix elements $M_{2,4}$ and $M_{4,6}$ are comparable to the activation gap, but the higher
transition $M_{2,6}$ is merely one tenth of them. So it is valid to neglect the mixing from higher LLs and restrict the perturbative approach to the first five LLs in
order to illustrate the result qualitatively.

We finally remark that the above computation is only valid in the perturbation sense. The Hartree-Fock ground state of $p$ filled LLs should
be calculated in a self-consistent way in order to give the accurate activation gap. However there is no such a self-consistent approach in FQH problems.
Nonetheless, our calculation is valid when the inter-CF LL transitions are small, as reflected in our work. It provides a qualitative criterion
for the robustness of activation gaps. In particular, in the computation of the optimal metric, we observe that when tuning the metric, the transition amplitude
goes from positive values to negative values. So at some point we can turn off the transition from, for example, $n=1$ to $n=3$ Landau levels. This means
for the optimal metric, the Hartree-Fock ansatz is self-consistent in the first few LLs.

\section{Conclusion}

In this paper, we build an anisotropic liquid state from the Hamiltonian point of view. The CF LLs and angular momenta can be combined
with a geometrical deformation of the Landau sites. A system under tilted magnetic field is studied to illustrate how the metric of this anisotropic liquid is deformed in response
to external anisotropy. We find that our results can be directly connected to the generalized pseudo-potential studies, as the inter-CF LL
transitions are proportional to the off-diagonal components of generalized pseudo-potentials. For $\nu=1/3$ and $\nu=2/5$ states, they almost have the
same optimal metric, defined as the minimal inter-LL transitions, implying the underlying off-diagonal pseudo-potentials. In the variation of the
metric, we also find the off-diagonal transitions turn from positive to negative, indicating that they actually vanish at some specific metric. So when we
take this metric, the CF description is almost self-consistent.

Furthermore, we find the activation gap is robust to the parallel magnetic field in the lowest Landau level, both for the flat metric and optimal metric.
Only in very extreme case $\tan\theta>3$ the activation gap is substantially modfied in wide quantum wells.
When the optimal metric is taken into account instead of the flat one, the activation gap is almost
unaffected by the tilt-induced anisotropy in the effective interaction potential,
exhibiting a great stability of the liquid phase, which is a reminiscent of the observation of enhanced FQHE in small tilting
angle \cite{PhysRevLett.108.196805}. Even in wide quantum wells with a strong inplane component of the magnetic field, the activation gap decreases at
$\nu=1/3$ and $2/5$ by only $10\dots 20\%$.

In our study, the metric is uniform in both space and time, treated as a parameter rather than a dynamical variable. From the point
view of a recently proposed bimetric theory \cite{PhysRevX.7.041032,liu2018geometric}, the metric computed here can be understood as the ambient metric.
How to build such a dynamical metric in the Hamiltonian theory will be an attractive extension of the construction here.

We also notice that in a recent work \cite{yang2018behavior}, the authors numerically calculated the magneto-roton gap for QH systems under tilted magnetic field.
There they show that the magneto-roton gap for filling $\nu=1/3$ closes for $\tan\theta>6$ and the structure factor exhibits charge density waves. We emphasize that this is
not in contradiction with respect to our results on the activation gap since the latter reflects the energy costed to activate a widely separated  quasi-particle/hole pair.
The anisotropy of the effective interaction potential is to great extent averaged in its computation. However, for wave-vector dependent quantities, such as the
magneto-roton excitation, the anisotropy manifests itself more directly. The magneto-roton gap is then likely to close before the charge gap, corresponding to a charge-density-wave
instability with a well-defined wave vector. Combined with our result, this may be a signature of nematic phase transition \cite{PhysRevB.96.035150}, where the
magneto-roton gap closes while the charged gap remains. The QH system under tilted magnetic field is thus a possible candidate to observe the nematic order.

\begin{acknowledgements}
We thank Nicolas Regnault for helpful discussions.
\end{acknowledgements}

\appendix*
\section{The effective interaction in the presence of a parallel magnetic field}\label{aptm}

As mentioned in the main text, the Hamiltonian of a quasi-2D system with a parabolic confinement potential in the $z$-direction is given by
\begin{equation}
H_{conf}=\sum_{i=x,y,z}\frac{(p_i+eA_i)}{2m}+\frac{m\Omega^2z^2}{2}.
\end{equation}

The gauge potential is chosen as $(0,-zB\tan\theta-xB,0)$ in the Landau gauge. Neglecting the Coulomb interaction, the Hamiltonian
is both quadratic in momenta and positions. It can be transformed into a canonical form by a coordinate transformation:
\begin{equation}
H_{conf}=\frac{p_\xi^2}{2m}+\frac{m\omega_-^2\xi^2}{2}+\frac{p_\zeta^2}{2m}+\frac{m\omega_+^2\zeta^2}{2}.
\end{equation}
The coordinates $\xi$ and $\zeta$ are related to the original coordinates by:
\begin{equation}
\left(\begin{array}{ccc}
\xi\\ \zeta\end{array}\right)=
\left(\begin{array}{ccc}
\cos\tilde\theta & -\sin\tilde\theta\\
\sin\tilde\theta & \cos\tilde\theta\end{array}\right)
\left(\begin{array}{ccc}
x \\ z\end{array}\right),
\end{equation}
where $\tan 2\tilde\theta=\tan 2\theta/(1-\tan^2\theta-\Omega^2/\omega_c^2)$. Now the Hamiltonian is in its canonical form and describes two decoupled
harmonic oscillators, one for the confined motion in the $z$-direction and another one for the cyclotron motion in the external
magnetic field. Their corresponding frequencies are given by
\begin{equation}
\frac{\omega^2_\pm}{\omega^2_c}=\frac{\lambda^2+1}{2}\pm\frac{\lambda^2-1}{2}\cos2\tilde\theta\mp\tan\theta\sin 2\tilde\theta,
\end{equation}
where $\lambda^2=(\tan^2\theta+\Omega^2/\omega_c^2)$. The two harmonic oscillators naturally define two length scales $l^2_\pm=(\omega_c/\omega_\pm)l^2$.

With such a decoupled Hamiltonian, the one-body states are denoted as $|N,n,m\rangle$, where the first index corresponding to the higher frequency
represents the level in the confining potential and the $n$ represents the cyclotron level, while the quantum number $m$ denotes the LL degeneracy (associated with the guding-center
coordinate). The lowest LL is naturally the $|0,0,m\rangle$.
To focus on the physics inside the lowest LL, the density operator is projected into this state:
\begin{equation}
\rho(q_x,q_y,q_z)=\sum_{m,m'}\langle0,0,m|e^{-i\mathbf \mathbf q\cdot\mathbf r}|0,0,m'\rangle c_m^{\dagger} c_{m'}.
\end{equation}
The indices of higher LLs depend on the exact strength of the confining potential: when $\Omega\to\infty$, the level $|0,n,m\rangle$
becomes the usual $n$-th LL for 2D electrons.

Inserting them into the interaction part of the Hamiltonian, after integrating out the $z$ coordinate, we obtain the effective two dimensional interaction:
\begin{align}
V_{\textrm{eff}}(\mathbf q)=&\int \frac{dq_z}{2\pi}\frac{4\pi e^2}{q^2}e^{-\frac{1}{2}\left[\frac{l_B^2 q_y^2\sin^2\tilde\theta}{l_+^2}+l^2_+(q_x\sin\tilde\theta-q_z\cos\tilde\theta)^2\right]}\nonumber\\
&\times e^{-\frac{1}{2}\left[\frac{l_B^2 q_y^2\cos^2\tilde\theta}{l_-^2}+l^2_-(q_x\cos\tilde\theta+q_z\sin\tilde\theta)^2\right]}\times\nonumber\\
&L^2_n\left[\frac{1}{2}\left(\frac{l_B^2 q_y^2\cos^2\tilde\theta}{l_-^2}+l^2_-(q_x\cos\tilde\theta+q_z\sin\tilde\theta)^2\right)\right].
\end{align}

In Ref. \cite{PhysRevLett.118.146403}, the authors integrate out the above expression in an arbitrary gauge and the effective potential
is expressed in terms of special functions. Here we stick to the integration expression in which a very practical Gaussian integral is available in numerical computations.

\bibliography{tilmag_actgp}

\begin{thebibliography}{24}%
\makeatletter
\providecommand \@ifxundefined [1]{%
 \@ifx{#1\undefined}
}%
\providecommand \@ifnum [1]{%
 \ifnum #1\expandafter \@firstoftwo
 \else \expandafter \@secondoftwo
 \fi
}%
\providecommand \@ifx [1]{%
 \ifx #1\expandafter \@firstoftwo
 \else \expandafter \@secondoftwo
 \fi
}%
\providecommand \natexlab [1]{#1}%
\providecommand \enquote  [1]{``#1''}%
\providecommand \bibnamefont  [1]{#1}%
\providecommand \bibfnamefont [1]{#1}%
\providecommand \citenamefont [1]{#1}%
\providecommand \href@noop [0]{\@secondoftwo}%
\providecommand \href [0]{\begingroup \@sanitize@url \@href}%
\providecommand \@href[1]{\@@startlink{#1}\@@href}%
\providecommand \@@href[1]{\endgroup#1\@@endlink}%
\providecommand \@sanitize@url [0]{\catcode `\\12\catcode `\$12\catcode
  `\&12\catcode `\#12\catcode `\^12\catcode `\_12\catcode `\%12\relax}%
\providecommand \@@startlink[1]{}%
\providecommand \@@endlink[0]{}%
\providecommand \url  [0]{\begingroup\@sanitize@url \@url }%
\providecommand \@url [1]{\endgroup\@href {#1}{\urlprefix }}%
\providecommand \urlprefix  [0]{URL }%
\providecommand \Eprint [0]{\href }%
\providecommand \doibase [0]{http://dx.doi.org/}%
\providecommand \selectlanguage [0]{\@gobble}%
\providecommand \bibinfo  [0]{\@secondoftwo}%
\providecommand \bibfield  [0]{\@secondoftwo}%
\providecommand \translation [1]{[#1]}%
\providecommand \BibitemOpen [0]{}%
\providecommand \bibitemStop [0]{}%
\providecommand \bibitemNoStop [0]{.\EOS\space}%
\providecommand \EOS [0]{\spacefactor3000\relax}%
\providecommand \BibitemShut  [1]{\csname bibitem#1\endcsname}%
\let\auto@bib@innerbib\@empty
%</preamble>
\bibitem [{\citenamefont {Wen}(2004)}]{wen2004quantum}%
  \BibitemOpen
  \bibfield  {author} {\bibinfo {author} {\bibfnamefont {X.}~\bibnamefont
  {Wen}},\ }\href {https://books.google.fr/books?id=RYESDAAAQBAJ} {\emph
  {\bibinfo {title} {Quantum Field Theory of Many-Body Systems: From the Origin
  of Sound to an Origin of Light and Electrons}}},\ Oxford Graduate Texts\
  (\bibinfo  {publisher} {OUP Oxford},\ \bibinfo {year} {2004})\BibitemShut
  {NoStop}%
\bibitem [{\citenamefont {Papi\ifmmode~\acute{c}\else
  \'{c}\fi{}}(2013)}]{PhysRevB.87.245315}%
  \BibitemOpen
  \bibfield  {author} {\bibinfo {author} {\bibfnamefont {Z.}~\bibnamefont
  {Papi\ifmmode~\acute{c}\else \'{c}\fi{}}},\ }\href {\doibase
  10.1103/PhysRevB.87.245315} {\bibfield  {journal} {\bibinfo  {journal} {Phys.
  Rev. B}\ }\textbf {\bibinfo {volume} {87}},\ \bibinfo {pages} {245315}
  (\bibinfo {year} {2013})}\BibitemShut {NoStop}%
\bibitem [{\citenamefont {Zhu}\ \emph {et~al.}(2017)\citenamefont {Zhu},
  \citenamefont {Sodemann}, \citenamefont {Sheng},\ and\ \citenamefont
  {Fu}}]{PhysRevB.95.201116}%
  \BibitemOpen
  \bibfield  {author} {\bibinfo {author} {\bibfnamefont {Z.}~\bibnamefont
  {Zhu}}, \bibinfo {author} {\bibfnamefont {I.}~\bibnamefont {Sodemann}},
  \bibinfo {author} {\bibfnamefont {D.~N.}\ \bibnamefont {Sheng}}, \ and\
  \bibinfo {author} {\bibfnamefont {L.}~\bibnamefont {Fu}},\ }\href {\doibase
  10.1103/PhysRevB.95.201116} {\bibfield  {journal} {\bibinfo  {journal} {Phys.
  Rev. B}\ }\textbf {\bibinfo {volume} {95}},\ \bibinfo {pages} {201116}
  (\bibinfo {year} {2017})}\BibitemShut {NoStop}%
\bibitem [{\citenamefont {Xia}\ \emph {et~al.}(2011)\citenamefont {Xia},
  \citenamefont {Eisenstein}, \citenamefont {Pfeiffer},\ and\ \citenamefont
  {West}}]{xia2011evidence}%
  \BibitemOpen
  \bibfield  {author} {\bibinfo {author} {\bibfnamefont {J.}~\bibnamefont
  {Xia}}, \bibinfo {author} {\bibfnamefont {J.}~\bibnamefont {Eisenstein}},
  \bibinfo {author} {\bibfnamefont {L.~N.}\ \bibnamefont {Pfeiffer}}, \ and\
  \bibinfo {author} {\bibfnamefont {K.~W.}\ \bibnamefont {West}},\ }\href@noop
  {} {\bibfield  {journal} {\bibinfo  {journal} {Nature Physics}\ }\textbf
  {\bibinfo {volume} {7}},\ \bibinfo {pages} {845} (\bibinfo {year}
  {2011})}\BibitemShut {NoStop}%
\bibitem [{\citenamefont {Liu}\ \emph {et~al.}(2012)\citenamefont {Liu},
  \citenamefont {Zhang}, \citenamefont {Tsui}, \citenamefont {Knez},
  \citenamefont {Levine}, \citenamefont {Du}, \citenamefont {Pfeiffer},\ and\
  \citenamefont {West}}]{PhysRevLett.108.196805}%
  \BibitemOpen
  \bibfield  {author} {\bibinfo {author} {\bibfnamefont {G.}~\bibnamefont
  {Liu}}, \bibinfo {author} {\bibfnamefont {C.}~\bibnamefont {Zhang}}, \bibinfo
  {author} {\bibfnamefont {D.~C.}\ \bibnamefont {Tsui}}, \bibinfo {author}
  {\bibfnamefont {I.}~\bibnamefont {Knez}}, \bibinfo {author} {\bibfnamefont
  {A.}~\bibnamefont {Levine}}, \bibinfo {author} {\bibfnamefont {R.~R.}\
  \bibnamefont {Du}}, \bibinfo {author} {\bibfnamefont {L.~N.}\ \bibnamefont
  {Pfeiffer}}, \ and\ \bibinfo {author} {\bibfnamefont {K.~W.}\ \bibnamefont
  {West}},\ }\href {\doibase 10.1103/PhysRevLett.108.196805} {\bibfield
  {journal} {\bibinfo  {journal} {Phys. Rev. Lett.}\ }\textbf {\bibinfo
  {volume} {108}},\ \bibinfo {pages} {196805} (\bibinfo {year}
  {2012})}\BibitemShut {NoStop}%
\bibitem [{\citenamefont {Fogler}\ \emph {et~al.}(1996)\citenamefont {Fogler},
  \citenamefont {Koulakov},\ and\ \citenamefont
  {Shklovskii}}]{PhysRevB.54.1853}%
  \BibitemOpen
  \bibfield  {author} {\bibinfo {author} {\bibfnamefont {M.~M.}\ \bibnamefont
  {Fogler}}, \bibinfo {author} {\bibfnamefont {A.~A.}\ \bibnamefont
  {Koulakov}}, \ and\ \bibinfo {author} {\bibfnamefont {B.~I.}\ \bibnamefont
  {Shklovskii}},\ }\href {\doibase 10.1103/PhysRevB.54.1853} {\bibfield
  {journal} {\bibinfo  {journal} {Phys. Rev. B}\ }\textbf {\bibinfo {volume}
  {54}},\ \bibinfo {pages} {1853} (\bibinfo {year} {1996})}\BibitemShut
  {NoStop}%
\bibitem [{\citenamefont {Moessner}\ and\ \citenamefont
  {Chalker}(1996)}]{PhysRevB.54.5006}%
  \BibitemOpen
  \bibfield  {author} {\bibinfo {author} {\bibfnamefont {R.}~\bibnamefont
  {Moessner}}\ and\ \bibinfo {author} {\bibfnamefont {J.~T.}\ \bibnamefont
  {Chalker}},\ }\href {\doibase 10.1103/PhysRevB.54.5006} {\bibfield  {journal}
  {\bibinfo  {journal} {Phys. Rev. B}\ }\textbf {\bibinfo {volume} {54}},\
  \bibinfo {pages} {5006} (\bibinfo {year} {1996})}\BibitemShut {NoStop}%
\bibitem [{\citenamefont {Maciejko}\ \emph {et~al.}(2013)\citenamefont
  {Maciejko}, \citenamefont {Hsu}, \citenamefont {Kivelson}, \citenamefont
  {Park},\ and\ \citenamefont {Sondhi}}]{PhysRevB.88.125137}%
  \BibitemOpen
  \bibfield  {author} {\bibinfo {author} {\bibfnamefont {J.}~\bibnamefont
  {Maciejko}}, \bibinfo {author} {\bibfnamefont {B.}~\bibnamefont {Hsu}},
  \bibinfo {author} {\bibfnamefont {S.~A.}\ \bibnamefont {Kivelson}}, \bibinfo
  {author} {\bibfnamefont {Y.}~\bibnamefont {Park}}, \ and\ \bibinfo {author}
  {\bibfnamefont {S.~L.}\ \bibnamefont {Sondhi}},\ }\href {\doibase
  10.1103/PhysRevB.88.125137} {\bibfield  {journal} {\bibinfo  {journal} {Phys.
  Rev. B}\ }\textbf {\bibinfo {volume} {88}},\ \bibinfo {pages} {125137}
  (\bibinfo {year} {2013})}\BibitemShut {NoStop}%
\bibitem [{\citenamefont {Jain}(1989)}]{PhysRevLett.63.199}%
  \BibitemOpen
  \bibfield  {author} {\bibinfo {author} {\bibfnamefont {J.~K.}\ \bibnamefont
  {Jain}},\ }\href {\doibase 10.1103/PhysRevLett.63.199} {\bibfield  {journal}
  {\bibinfo  {journal} {Phys. Rev. Lett.}\ }\textbf {\bibinfo {volume} {63}},\
  \bibinfo {pages} {199} (\bibinfo {year} {1989})}\BibitemShut {NoStop}%
\bibitem [{\citenamefont {Jain}(1990)}]{PhysRevB.41.7653}%
  \BibitemOpen
  \bibfield  {author} {\bibinfo {author} {\bibfnamefont {J.~K.}\ \bibnamefont
  {Jain}},\ }\href {\doibase 10.1103/PhysRevB.41.7653} {\bibfield  {journal}
  {\bibinfo  {journal} {Phys. Rev. B}\ }\textbf {\bibinfo {volume} {41}},\
  \bibinfo {pages} {7653} (\bibinfo {year} {1990})}\BibitemShut {NoStop}%
\bibitem [{\citenamefont {Lopez}\ and\ \citenamefont
  {Fradkin}(1991)}]{PhysRevB.44.5246}%
  \BibitemOpen
  \bibfield  {author} {\bibinfo {author} {\bibfnamefont {A.}~\bibnamefont
  {Lopez}}\ and\ \bibinfo {author} {\bibfnamefont {E.}~\bibnamefont
  {Fradkin}},\ }\href {\doibase 10.1103/PhysRevB.44.5246} {\bibfield  {journal}
  {\bibinfo  {journal} {Phys. Rev. B}\ }\textbf {\bibinfo {volume} {44}},\
  \bibinfo {pages} {5246} (\bibinfo {year} {1991})}\BibitemShut {NoStop}%
\bibitem [{\citenamefont {Murthy}\ and\ \citenamefont
  {Shankar}(2003)}]{RevModPhys.75.1101}%
  \BibitemOpen
  \bibfield  {author} {\bibinfo {author} {\bibfnamefont {G.}~\bibnamefont
  {Murthy}}\ and\ \bibinfo {author} {\bibfnamefont {R.}~\bibnamefont
  {Shankar}},\ }\href {\doibase 10.1103/RevModPhys.75.1101} {\bibfield
  {journal} {\bibinfo  {journal} {Rev. Mod. Phys.}\ }\textbf {\bibinfo {volume}
  {75}},\ \bibinfo {pages} {1101} (\bibinfo {year} {2003})}\BibitemShut
  {NoStop}%
\bibitem [{\citenamefont {Yang}\ \emph
  {et~al.}(2017{\natexlab{a}})\citenamefont {Yang}, \citenamefont {Lee},
  \citenamefont {Zhang},\ and\ \citenamefont {Hu}}]{PhysRevB.96.195140}%
  \BibitemOpen
  \bibfield  {author} {\bibinfo {author} {\bibfnamefont {B.}~\bibnamefont
  {Yang}}, \bibinfo {author} {\bibfnamefont {C.~H.}\ \bibnamefont {Lee}},
  \bibinfo {author} {\bibfnamefont {C.}~\bibnamefont {Zhang}}, \ and\ \bibinfo
  {author} {\bibfnamefont {Z.-X.}\ \bibnamefont {Hu}},\ }\href {\doibase
  10.1103/PhysRevB.96.195140} {\bibfield  {journal} {\bibinfo  {journal} {Phys.
  Rev. B}\ }\textbf {\bibinfo {volume} {96}},\ \bibinfo {pages} {195140}
  (\bibinfo {year} {2017}{\natexlab{a}})}\BibitemShut {NoStop}%
\bibitem [{\citenamefont {Yang}\ \emph {et~al.}(2018)\citenamefont {Yang},
  \citenamefont {Li},\ and\ \citenamefont {Hu}}]{yang2018behavior}%
  \BibitemOpen
  \bibfield  {author} {\bibinfo {author} {\bibfnamefont {L.-P.}\ \bibnamefont
  {Yang}}, \bibinfo {author} {\bibfnamefont {Q.}~\bibnamefont {Li}}, \ and\
  \bibinfo {author} {\bibfnamefont {Z.-X.}\ \bibnamefont {Hu}},\ }\href@noop {}
  {\bibfield  {journal} {\bibinfo  {journal} {arXiv preprint arXiv:1805.00805}\
  } (\bibinfo {year} {2018})}\BibitemShut {NoStop}%
\bibitem [{\citenamefont {Haldane}(2011)}]{PhysRevLett.107.116801}%
  \BibitemOpen
  \bibfield  {author} {\bibinfo {author} {\bibfnamefont {F.~D.~M.}\
  \bibnamefont {Haldane}},\ }\href {\doibase 10.1103/PhysRevLett.107.116801}
  {\bibfield  {journal} {\bibinfo  {journal} {Phys. Rev. Lett.}\ }\textbf
  {\bibinfo {volume} {107}},\ \bibinfo {pages} {116801} (\bibinfo {year}
  {2011})}\BibitemShut {NoStop}%
\bibitem [{\citenamefont {Haldane}(1983)}]{PhysRevLett.51.605}%
  \BibitemOpen
  \bibfield  {author} {\bibinfo {author} {\bibfnamefont {F.~D.~M.}\
  \bibnamefont {Haldane}},\ }\href {\doibase 10.1103/PhysRevLett.51.605}
  {\bibfield  {journal} {\bibinfo  {journal} {Phys. Rev. Lett.}\ }\textbf
  {\bibinfo {volume} {51}},\ \bibinfo {pages} {605} (\bibinfo {year}
  {1983})}\BibitemShut {NoStop}%
\bibitem [{\citenamefont {Yang}\ \emph
  {et~al.}(2017{\natexlab{b}})\citenamefont {Yang}, \citenamefont {Hu},
  \citenamefont {Lee},\ and\ \citenamefont {Papi\ifmmode~\acute{c}\else
  \'{c}\fi{}}}]{PhysRevLett.118.146403}%
  \BibitemOpen
  \bibfield  {author} {\bibinfo {author} {\bibfnamefont {B.}~\bibnamefont
  {Yang}}, \bibinfo {author} {\bibfnamefont {Z.-X.}\ \bibnamefont {Hu}},
  \bibinfo {author} {\bibfnamefont {C.~H.}\ \bibnamefont {Lee}}, \ and\
  \bibinfo {author} {\bibfnamefont {Z.}~\bibnamefont
  {Papi\ifmmode~\acute{c}\else \'{c}\fi{}}},\ }\href {\doibase
  10.1103/PhysRevLett.118.146403} {\bibfield  {journal} {\bibinfo  {journal}
  {Phys. Rev. Lett.}\ }\textbf {\bibinfo {volume} {118}},\ \bibinfo {pages}
  {146403} (\bibinfo {year} {2017}{\natexlab{b}})}\BibitemShut {NoStop}%
\bibitem [{\citenamefont {Hu}\ \emph {et~al.}(2018)\citenamefont {Hu},
  \citenamefont {Li}, \citenamefont {Yang}, \citenamefont {Yang}, \citenamefont
  {Jiang}, \citenamefont {Qiu},\ and\ \citenamefont
  {Yang}}]{PhysRevB.97.035140}%
  \BibitemOpen
  \bibfield  {author} {\bibinfo {author} {\bibfnamefont {Z.-X.}\ \bibnamefont
  {Hu}}, \bibinfo {author} {\bibfnamefont {Q.}~\bibnamefont {Li}}, \bibinfo
  {author} {\bibfnamefont {L.-P.}\ \bibnamefont {Yang}}, \bibinfo {author}
  {\bibfnamefont {W.-Q.}\ \bibnamefont {Yang}}, \bibinfo {author}
  {\bibfnamefont {N.}~\bibnamefont {Jiang}}, \bibinfo {author} {\bibfnamefont
  {R.-Z.}\ \bibnamefont {Qiu}}, \ and\ \bibinfo {author} {\bibfnamefont
  {B.}~\bibnamefont {Yang}},\ }\href {\doibase 10.1103/PhysRevB.97.035140}
  {\bibfield  {journal} {\bibinfo  {journal} {Phys. Rev. B}\ }\textbf {\bibinfo
  {volume} {97}},\ \bibinfo {pages} {035140} (\bibinfo {year}
  {2018})}\BibitemShut {NoStop}%
\bibitem [{\citenamefont {Qiu}\ \emph {et~al.}(2012)\citenamefont {Qiu},
  \citenamefont {Haldane}, \citenamefont {Wan}, \citenamefont {Yang},\ and\
  \citenamefont {Yi}}]{PhysRevB.85.115308}%
  \BibitemOpen
  \bibfield  {author} {\bibinfo {author} {\bibfnamefont {R.-Z.}\ \bibnamefont
  {Qiu}}, \bibinfo {author} {\bibfnamefont {F.~D.~M.}\ \bibnamefont {Haldane}},
  \bibinfo {author} {\bibfnamefont {X.}~\bibnamefont {Wan}}, \bibinfo {author}
  {\bibfnamefont {K.}~\bibnamefont {Yang}}, \ and\ \bibinfo {author}
  {\bibfnamefont {S.}~\bibnamefont {Yi}},\ }\href {\doibase
  10.1103/PhysRevB.85.115308} {\bibfield  {journal} {\bibinfo  {journal} {Phys.
  Rev. B}\ }\textbf {\bibinfo {volume} {85}},\ \bibinfo {pages} {115308}
  (\bibinfo {year} {2012})}\BibitemShut {NoStop}%
\bibitem [{\citenamefont {Murthy}(1999)}]{PhysRevB.60.13702}%
  \BibitemOpen
  \bibfield  {author} {\bibinfo {author} {\bibfnamefont {G.}~\bibnamefont
  {Murthy}},\ }\href {\doibase 10.1103/PhysRevB.60.13702} {\bibfield  {journal}
  {\bibinfo  {journal} {Phys. Rev. B}\ }\textbf {\bibinfo {volume} {60}},\
  \bibinfo {pages} {13702} (\bibinfo {year} {1999})}\BibitemShut {NoStop}%
\bibitem [{\citenamefont {Stanescu}\ \emph {et~al.}(2000)\citenamefont
  {Stanescu}, \citenamefont {Martin},\ and\ \citenamefont
  {Phillips}}]{PhysRevLett.84.1288}%
  \BibitemOpen
  \bibfield  {author} {\bibinfo {author} {\bibfnamefont {T.~D.}\ \bibnamefont
  {Stanescu}}, \bibinfo {author} {\bibfnamefont {I.}~\bibnamefont {Martin}}, \
  and\ \bibinfo {author} {\bibfnamefont {P.}~\bibnamefont {Phillips}},\ }\href
  {\doibase 10.1103/PhysRevLett.84.1288} {\bibfield  {journal} {\bibinfo
  {journal} {Phys. Rev. Lett.}\ }\textbf {\bibinfo {volume} {84}},\ \bibinfo
  {pages} {1288} (\bibinfo {year} {2000})}\BibitemShut {NoStop}%
\bibitem [{\citenamefont {Gromov}\ and\ \citenamefont
  {Son}(2017)}]{PhysRevX.7.041032}%
  \BibitemOpen
  \bibfield  {author} {\bibinfo {author} {\bibfnamefont {A.}~\bibnamefont
  {Gromov}}\ and\ \bibinfo {author} {\bibfnamefont {D.~T.}\ \bibnamefont
  {Son}},\ }\href {\doibase 10.1103/PhysRevX.7.041032} {\bibfield  {journal}
  {\bibinfo  {journal} {Phys. Rev. X}\ }\textbf {\bibinfo {volume} {7}},\
  \bibinfo {pages} {041032} (\bibinfo {year} {2017})}\BibitemShut {NoStop}%
\bibitem [{\citenamefont {Liu}\ \emph {et~al.}(2018)\citenamefont {Liu},
  \citenamefont {Gromov},\ and\ \citenamefont {Papi{\'c}}}]{liu2018geometric}%
  \BibitemOpen
  \bibfield  {author} {\bibinfo {author} {\bibfnamefont {Z.}~\bibnamefont
  {Liu}}, \bibinfo {author} {\bibfnamefont {A.}~\bibnamefont {Gromov}}, \ and\
  \bibinfo {author} {\bibfnamefont {Z.}~\bibnamefont {Papi{\'c}}},\ }\href@noop
  {} {\bibfield  {journal} {\bibinfo  {journal} {arXiv preprint
  arXiv:1803.00030}\ } (\bibinfo {year} {2018})}\BibitemShut {NoStop}%
\bibitem [{\citenamefont {Regnault}\ \emph {et~al.}(2017)\citenamefont
  {Regnault}, \citenamefont {Maciejko}, \citenamefont {Kivelson},\ and\
  \citenamefont {Sondhi}}]{PhysRevB.96.035150}%
  \BibitemOpen
  \bibfield  {author} {\bibinfo {author} {\bibfnamefont {N.}~\bibnamefont
  {Regnault}}, \bibinfo {author} {\bibfnamefont {J.}~\bibnamefont {Maciejko}},
  \bibinfo {author} {\bibfnamefont {S.~A.}\ \bibnamefont {Kivelson}}, \ and\
  \bibinfo {author} {\bibfnamefont {S.~L.}\ \bibnamefont {Sondhi}},\ }\href
  {\doibase 10.1103/PhysRevB.96.035150} {\bibfield  {journal} {\bibinfo
  {journal} {Phys. Rev. B}\ }\textbf {\bibinfo {volume} {96}},\ \bibinfo
  {pages} {035150} (\bibinfo {year} {2017})}\BibitemShut {NoStop}%
\end{thebibliography}%

\end{document}